\documentclass[a4paper,12pt]{article}
\pdfoutput=1
\usepackage{graphicx} 
\usepackage{preamble}
\usepackage{colonequals}
\usepackage{simplewick}
\usepackage{feynmp-auto}
\usepackage{authblk}
\usepackage[sorting=none,style=numeric-comp,maxbibnames=10]{biblatex}
\usepackage[normalem]{ulem}

\newcommand{\hq}{\hat q}
\newcommand{\eqgraph}[3]{%
  \begin{gathered}
  \raisebox{0pt}[\dimexpr\height+#1][\dimexpr\depth+#2]{\ignorespaces#3\unskip}%
  \end{gathered}
}

\makeatletter
\def\blfootnote{\gdef\@thefnmark{}\@footnotetext}
\makeatother

\title{A Cordial Introduction to Double Scaled SYK}

\author[1]{Micha Berkooz} 
\author[2,3]{Ohad Mamroud}
\affil[1]{\normalsize \it Department of Particle Physics and Astrophysics, Weizmann Institute of Science, Rehovot 7610001, Israel}
\affil[2]{\normalsize  \it SISSA, via Bonomea 265, 34136 Trieste, Italy}
\affil[3]{\normalsize \it INFN, Sezione di Trieste, Via Valerio 2, 34127 Trieste, Italy}

\nocite{kitaev2015simple, Maldacena:2016hyu,  sachdev1993,sachdev2010,Chowdhury:2021qpy,Maldacena:2016upp,Erdos,Berkooz:2018qkz,Berkooz:2018jqr}
\begin{document}

\maketitle

\begin{abstract}
We review recent progress regarding the double scaled Sachdev-Ye-Kitaev model and other $p$-local quantum mechanical random Hamiltonians. These models exhibit an expansion using chord diagrams, which can be solved by combinatorial methods. We describe exact results in these models, including their spectrum, correlation functions, and Lyapunov exponent. In a certain limit, these techniques manifest the relation to the Schwarzian quantum mechanics, a theory of quantum gravity in $AdS_2$. More generally, the theory is controlled by a rigid algebraic structure of a quantum group, suggesting a theory of quantum gravity on non-commutative $q$-deformed $AdS_2$. We conclude with discussion of related universality classes, and survey some of the current research directions.
\end{abstract}

\tableofcontents

\section{Introduction}

The Sachdev-Ye-Kitaev (SYK) model \cite{kitaev2015simple, Maldacena:2016hyu,  sachdev1993, sachdev2010} is a quantum mechanical model of $N$ Majorana fermions with all-to-all $p$-local interactions. The model is quantum chaotic, as indicated by level repulsion \cite{you2017sachdev, garcia2016,Cotler:2016fpe} and its maximal chaos exponent \cite{kitaev2015simple, Maldacena:2016hyu, Maldacena:2015waa}. Variants of the SYK model appeared early in nuclear physics \cite{french1970,bohigas1971}  but it became particularly useful in condensed matter physics  as a model for strange metals \cite{sachdev1993,sachdev2010,Chowdhury:2021qpy}. Its maximal chaos exponent made it important as a controllable model of holography \cite{kitaev2015simple, Maldacena:2016upp, Cotler:2016fpe, Saad:2018bqo, Maldacena:2018lmt, Goel:2018ubv, Jensen:2016pah, Polchinski:2016xgd, sachdev2010} (and many others). 

This review concerns the double scaled SYK model (DS-SYK), where instead of a standard large $N$ limit one takes a \emph{double scaling limit}, in which the length of the interaction also scales with $N$. The model then describes a generalization of the Schwarzian theory which holds at all energy scales. We review how this structure naturally emerges from DS-SYK via the diagrammatics of \emph{chord diagrams}, point at its underlying algebraic structure and the role of non-commutative geometry, and comment on possible holographic realizations.

Actually, much of what we will say also applies to $p$-local systems in general, encompassing a rich \emph{universality class}. But let us for now focus on the Majorana SYK model: 
\begin{equation}
    \label{eq:SYK Hamiltonian}
    \ H = i^{p/2} \sum_{1 \le i_1 < i_2 <\cdots < i_p \le N} J_{i_1 \cdots i_p} \psi_{i_1} \cdots \psi_{i_p}  \, , \qquad \{\psi_i, \psi_j\} = 2\delta_{ij}\, \qquad i,j=1,\dots,N \,.
\end{equation}
$H$ is a $p$-local Hamiltonian\footnote{For finite $p$ the SYK model is special in the sense that it has a non-Fermi-liquid phase at low energies rather than a spin glass phase.}, as it is a sum of terms, each acting on $p$ out of the $N$ available degrees of freedom.
At large $N$ the theory is in fact nearly conformal in the IR, which matches the symmetry of JT gravity on near-$AdS_2$ and results in the Schwarzian theory.

We will be interested in a different limit of the model -- the \emph{double scaling} limit
\begin{equation}
p,\, N\rightarrow \infty, \quad \lambda\propto p^2/N\ fixed\,.
\end{equation}
Throughout the review we will take the strict limit without any $1/N$ corrections. This limit has advantages for the study of the thermodynamics and chaos properties of the SYK model, and $p$-local systems in general; for holography; and for random matrix theory (RMT).

\paragraph{For $p$-local systems:}

The model can be solved efficiently using combinatorial tools. We can write closed expressions for the free energy and correlators at all energy scales\footnote{At times that remain finite as we take $N\rightarrow\infty$.}, rather than just at low energies or having to solve additional differential equations (sections \ref{sec:HamChords} and \ref{sec:four_point_func}). Furthermore, the rules for this limit are the same for a large class of $p$-local systems, the only difference being the precise relation between $\lambda$ and the underlying microscopic model, exhibiting the universal nature of the construction (sections \ref{sec:HamChords} and \ref{sec:Generalizations}). 

In addition, there is a natural set of local observables which are fairly universal across many $p$-local models (Section \ref{sec:HamChords}). Together with the Hamiltonian, they form multi-matrix models where non-trivial correlation appear just because of the $p$-local structure, rather then due to direct couplings in a potential. In particular one can compute their out-of-time-order correlators exactly for cases with both maximal and submaximal chaos (Section \ref{sec:four_point_func}). 

It is possible to generalize the theory to sums of $p$-local Hamiltonians with different spectral properties, and solve it in some limits. This allows the study of RG flows, chaos-integrability transitions, and quantum critical points (Section \ref{sec:MltChrd}). 

\paragraph{For Holography:}

An interesting feature of the DS-SYK model is that it makes the SYK/2D gravity duality manifest. The combinatorial tools that we will present start from the microscopic data and, after ensemble averaging, produce the bulk gravitational Hilbert space and Hamiltonian. For finite $\lambda$ it could be interpreted as having a minimal length in the bulk, so we can think about it as a full model of quantum gravity. 

When $\lambda \to 0$ the low energy spectrum reproduces the Schwarzian theory. Thus in the IR the theory is holographically dual to JT gravity in near-$AdS_2$. But in DS-SYK we can also access the rest of the spectrum. It keeps exhibiting certain holographic features, the only difference being that the temperature is replaced by a ``fake'' temperature (Section~\ref{sec:semi-classics}). The chords provide an underlying symmetry that enforces these properties (Section~\ref{sec:chord algebra}).

At finite $\lambda$ the theory is ``algebraically rigid'', as correlation functions, and the Lyapunov exponent, are given by group theoretic objects related to the quantum group $SU_{\hq}(1,1)$ which is a $q$-deformation of $SU(1,1)$ (Section \ref{sec:four_point_func}). This algebraic structure seems to be reflected in the bulk by changing the familiar $AdS_2$ into a non-commutative space (Section~\ref{sec:non-commutative}).

\paragraph{For RMT:} Following 't Hooft we are used to 2D surfaces arising from planar diagrams and large $N$ limits. The way it is done is similar to free probability theory and exploits the features of the standard large $N$ Wigner-Dyson $\beta$-ensembles, where the Hamiltonians $H$ are drawn from a distribution function of the form $\exp(-\dim({\cal H}) V(H))$ where $V$ is invariant under $SU(\dim {\cal H})$ or other such groups.
The $p$-local models that we discuss here do not fall into this class, yet they still give rise to 2D surfaces. They are solvable using manageable combinatorial tools, a fact that was already appreciated in the mathematics community \cite{Speicher:1993kt,speicher1991example,speicher1997q}, with the structure of chords (that we describe below) allowing for an elegant way of going beyond free probability theory.\newline

Finally, Section \ref{sec:Frntrs} mentions some open problems: a gravitational path integral formulation, the relation to de-Sitter space (and type II$_1$ von-Neumann algebras), multi-trace correlators and Euclidean wormholes, how local geometry arises in the bulk and the relation to operator complexity.

All of of these topics have been developed in quite a few papers. Above we only indicated where the different topics appear in the review, and more detailed references appear in each section. It is unlikely that we have managed to do justice to the different topics, or to the publications and authors who worked them out and we apologize for this.  


\section{Chord Diagrams and the partition function}\label{sec:HamChords}

The main technical tool used to solve the model are \emph{chord diagrams}:

\paragraph{Chord diagram:} Consider $k$ (even) dots on a circle. A chord diagram is a pairing of these points. We denote it by connecting each pair of points by a chord, see \autoref{fig:chord diagram}.

\medskip
These appear quite naturally when evaluating the moments $\Tr(H_{rand}^k)$ of random Hamiltonians of the form
\begin{equation}\label{eq:HRandA}
H_{rand}=\sum_\alpha J_\alpha {\cal O}_\alpha \,,
\end{equation}
where $J_\alpha$ are Gaussian random variables, and ${\cal O}_\alpha$ is any set of operators acting on the Hilbert space. When evaluating the ensemble average over the couplings $\langle\Tr(H^{k}_{rand})\rangle_J$, the chord diagrams just encode the Gaussian averages of the $J_\alpha$ via Wick's theorem. The choice of ${\cal O}_\alpha$ then determines the weight of the chord diagram, and changes from model to model. Some choices of the weights give rise to universality classes --  the ordinary Wigner-Dyson (Gaussian) $\beta$-ensembles are just those with only non-intersecting chords, while our universality class is that in which there is a fixed weight for each chord intersection \cite{Speicher:1993kt,speicher1991example,speicher1997q}. Let us see how this works for the DS-SYK model.

\subsection{Chord diagrams}\label{sec:ChrdDiag}

The SYK model is a microscopic model of $N$ Majorana fermions $\psi_i$, $\{\psi_i, \psi_j\} = 2\delta_{ij}$, with
\begin{equation}
    \label{eq:SYK Hamiltonian}
    H = i^{p/2} \sum_{i_1 < i_2 <\cdots < i_p = 1}^{N} J_{i_1 \cdots i_p} \psi_{i_1} \cdots \psi_{i_p} \equiv i^{p/2} \sum_I J_I \Psi_I \,,
\end{equation}
where the $J$'s are random couplings drawn from a Gaussian ensemble. $I$'s denote ordered sets of $p$ indices, $\{i_1, \cdots, i_p\}$, and $\Psi_I$ denotes the product of $p$ fermions with indices in $I$. At large $N$ the model is self-averaging, allowing us to study annealed averages over the couplings, denoted by $\langle \cdot \rangle_J$. The statistics of the couplings $J$ are
\begin{equation}
    \label{eq:cJ bJ normal}
    \left\langle J_I \right\rangle_J = 0 \,,\qquad \left\langle J_I J_K\right\rangle_J = \binom{N}{p}^{-1} \cJ^2 \delta_{IK} = \frac{N}{2p^2}\binom{N}{p}^{-1} \bJ^2 \delta_{IK}\,,
\end{equation}
where both the $\cJ$ \cite{Berkooz:2018jqr, Berkooz:2018qkz} and the $\bJ$ convention \cite{Maldacena:2016hyu, Lin:2022rbf, Lin:2023trc} exist in the literature, and are related by $\bJ^2 = \frac{2p^2}{N}\cJ^2$. We also choose a normalized\footnote{
\label{foo: trace norm}One could also use a physical convention where the trace of the identity is the size of the Hilbert space, $\tr(\1) = 2^{N/2}$. The extra factor simply multiplies quantities such as the partition function and adds a positive constant $\frac{N}{2}\log(2)$ to any computed entropy. Our choice to normalize the trace allows it to be well defined in the double scaling limit, but means that the entropies computed using it are always relative to the maximally mixed state. This is related to the appearance of type II$_1$ von Neumann algebra for the observables in the model, as mentioned in Section~\ref{sec:de Sitter}.} 
trace, $\Tr(\1) = 1$. These choices imply $\left\langle \Tr\left(H^2\right)\right\rangle_J = \cJ^2$. From now on we set $\cJ = 1$, and restore it by dimensional analysis when needed. We will be interested in the {\it double scaling limit}\footnote{As pointed out in \cite{Cotler:2016fpe}, it marks the boundary between $p$-local behavior for smaller $p$ and and nonlocal behavior for larger.}
\begin{equation}\label{eq:DblSclng}
    N, p \to \infty\,, \qquad \lambda \equiv 2p^2/N \quad fixed.
\end{equation}

\begin{figure}[t]
    \centering
    \begin{subfigure}[h]{0.45\textwidth}
    \centering
    \includegraphics[width=0.5\textwidth]{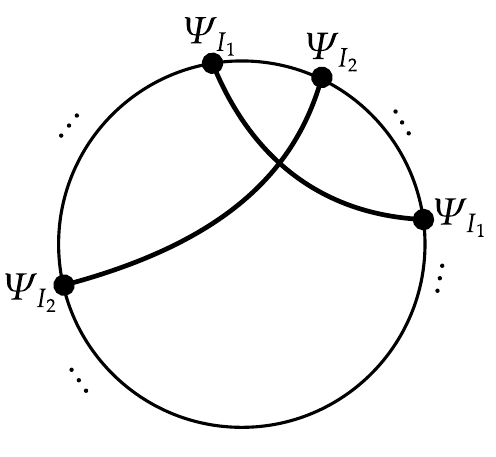}
    \end{subfigure}
    \qquad
    \begin{subfigure}[h]{0.45\textwidth}
    \centering
    \includegraphics[width=0.5\textwidth]{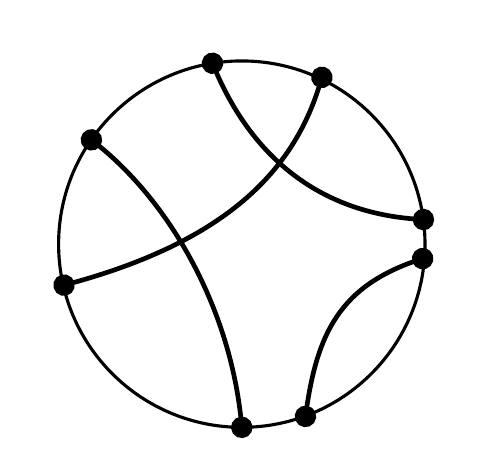}
    \end{subfigure}
    \caption{The chord diagram for \protect\eqref{eq:Moment contractions} (left) and for ${\protect\wick{\Tr \left(\c4{\Psi}_{I_1} \c3{\Psi}_{I_2} \c4{\Psi}_{I_1} \c1{\Psi}_{I_3} \c1{\Psi}_{I_3} \c2{\Psi}_{I_4} \c3{\Psi}_{I_2} \c2{\Psi}_{I_4} \right)}}$ (right). Here we also highlighed the nodes at the ends of each chord, but we will suppress them later.}
    \label{fig:chord diagram}
\end{figure}

We'll start by analyzing the partition function of the model by expanding it into moments,
\begin{align}
\label{eq:partition func into moments}
    Z\left(\beta\right) &= \left\langle \Tr \left(e^{-\beta H}\right) \right\rangle_J 
 = \sum_{k=0}^{\infty}\frac{\left(-\beta\right)^{k}}{k!}m_k \,, \\
 m_k &= \left\langle \Tr\left(H^{k}\right)\right\rangle_J = i^{kp/2}\sum_{I_{1},\dots,I_{k}}\left\langle J_{I_{1}}\cdots J_{I_{k}}\right\rangle_J \Tr\left(\Psi_{I_{1}}\cdots\Psi_{I_{k}}\right)\,.
\end{align}
Since the $J$'s are Gaussian variables we use Wick's theorem for the ensemble average. \eqref{eq:cJ bJ normal} implies that the indices of pairs of $\Psi$'s are identified, which we also represent as contractions,
\begin{equation}
\label{eq:Moment contractions}
    m_k = i^{kp/2}\sum_{I_{1},\dots,I_{k}}\binom{N}{p}^{-k/2} \sum_{I_1, \cdots, I_{k/2}} \sum_{\text{Wick cont.}} \Tr\left(\wick{\c3\Psi_{I_1} \c2\Psi_{I_2}\cdots\c3\Psi_{I_1}\cdots \c2\Psi_{I_2}\cdots} \cdots \right) \,.
\end{equation}
As the trace is cyclic, we can represent these contractions by a \emph{chord diagram} as in \autoref{fig:chord diagram}. The sum over all Wick contractions becomes a sum over all chord diagrams. 

Next we evaluate the trace with the prescribed ordering of the $\Psi_I$'s. We permute the $\Psi_I$'s until two with the same index set $I$ are adjacent, and then annihilate them as $\Psi_I^2\propto \1$. Commuting $\Psi_{I}$ and $\Psi_{K}$ introduces a sign that depends on the number of indices in common to $I$ and $K$, $(-1)^{\left|I \cap K\right|}$.
In the double scaling limit $\left|I\cap K\right|$ is Poisson distributed\footnote{
\label{foo: Poisson dist}
Overall there are $\binom{N}{p}^2$ ways of choosing the two lists. Sets with $m$ intersecting indices can be found by choosing the $p$ indices in $I$, then from them picking $m$ indices to $K$, and finally choosing the other $p-m$ indices in $K$ such that they are not in $I$. The probability $P(m)$ of having exactly $m$ intersecting indices is
\begin{equation}
    \frac{\binom{N}{p} \binom{p}{m} \binom{N-p}{p-m}}{\binom{N}{p}^2} = \frac{1}{m!}\left(\frac{p!}{(p-m)!}\right)^2 \frac{(N-p)!}{N!}\frac{(N-p)!}{(N-2p+m)!} = \frac{1}{m!}\left(\frac{p^2}{N}\right)^m \prod_{j=0}^p \frac{N-p-j}{N-j}\prod_{\ell=0}^m \frac{(1-\frac{\ell}{p})^2}{1-\frac{2p-\ell}{N}} \,,
\end{equation}
assuming $m$ is finite, taking the double scaling limit allows us to neglect the second product and simplify the first, such that the probability becomes
\begin{equation}
    P(m) = \frac{1}{m!}\left(\frac{p^2}{N}\right)^m \left(1-\frac{p}{N}\right)^p \rightarrow \frac{1}{m!}\left(\frac{p^2}{m}\right)^m e^{-p^2/N} \,.
\end{equation}} 
with mean $\frac{p^{2}}{N}$ since $I$ and $K$ are two independent lists of numbers of length $p$, drawn from $\{1,\cdots,N\}$ \cite{Erdos, Cotler:2016fpe}, while intersections of three or more index sets are empty with probability 1. 
The exchange of each $\Psi_I$ and $\Psi_K$ -- each intersection of chords in the diagram -- therefore gives a factor of $\langle (-1)^{|I\cap K|} \rangle_{Poisson}$ \cite{Cotler:2016fpe,Berkooz:2018jqr}
\begin{equation} 
    \sum_{m=0}^{\infty}\frac{\left(p^{2}/N\right)^{m}}{m!}e^{-p^{2}/N}\left(-1\right)^{m} = e^{-\lambda}\equiv q  \,.
\end{equation}
Summing over all chord diagrams, we find that each moment gives\footnote{In the $\bJ$ normalization of  \eqref{eq:cJ bJ normal} there is an additional factor of $\left(\frac{\bJ}{\sqrt{\lambda}}\right)^k$ appearing in the moment $m_k$.}
\begin{equation}
\label{eq:MmntsChrds}
m_{k}=\sum_{\substack{\text{chord diagrams}\\ \text{with $k/2$ chords}}}q^{\,\text{no. of intersections}} \,.
\end{equation}
This expression is valid when $k$ is held finite\footnote{Perhaps it can be extended to $k\sim N^\alpha$, $\alpha$ positive, small enough.} as $N\rightarrow\infty$.

Formally\footnote{The action is not bounded from below, and so the correct integration contour for $g$ is unclear. Similar action can be found when $q \to 1$ by coarse graining chord diagrams \cite{Berkooz:2024evs}, where the integral is better defined.}, one can write a generating function for chord diagrams\footnote{In SYK, the expression can be directly derived from the partition function of the underlying fermionic model by the $G\Sigma$ formalism, which re-writes the action in terms of the fermion two point function, $G(\tau_1,\tau_2) = \frac{1}{N}\sum_i \psi_i(\tau_1)\psi_2(\tau_2)$. In the double scaling limit ${G(\tau_1,\tau_2) = \sign(\tau_1 - \tau_2) e^{g(\tau_1,\tau_2)/p}}$, $\Sigma(\tau_1,\tau_2) = \sigma(\tau_1,\tau_2)/p$, and integrating out the latter gives \eqref{eq:GSigma double scaling} \cite{Cotler:2016fpe, Stanford-talk-kitp} (see also Section 4.1 of \cite{Goel:2023svz}). Yet, the generating function holds for any model described by chord diagrams, even without direct $G\Sigma$ like derivation.} \cite{Cotler:2016fpe, Lin:2023trc, Stanford-talk-kitp},
\begin{equation}
\label{eq:GSigma double scaling}
Z(\beta) = \int \cD g \exp\left[-\frac{1}{\lambda}\int_0^\beta d\tau_1 \int_0^{\tau_1} d\tau_2 \left[\frac{1}{4} \partial_1 g(\tau_1,\tau_2) \partial_2 g(\tau_1,\tau_2) - \bJ^2 e^{g(\tau_1,\tau_2)} \right]\right]\,,
\end{equation}
The $k$-th order in the perturbative expansion in $\bJ$  reproduces the contribution of the $k$-th moment to the partition function of temperature $\beta$, see Appendix H of \cite{Lin:2023trc} for details.


\subsection{The transfer matrix}\label{sec:TrnsfrMtrx}

\begin{figure}
    \centering
    \begin{subfigure}[t]{0.25\textwidth}
    \centering
    \includegraphics[width=1\textwidth]{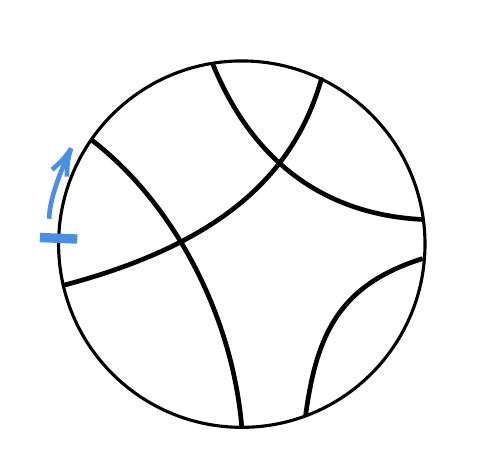}    
    \end{subfigure}
    \qquad
    \begin{subfigure}[t]{0.5\textwidth}
    \centering
    \includegraphics[width=1\textwidth]{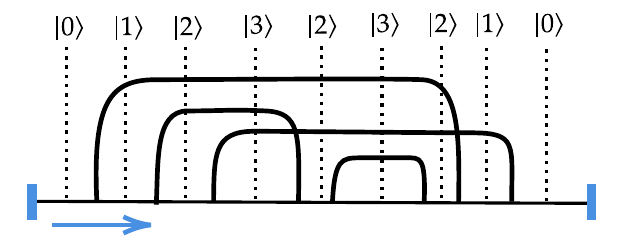} 
    \end{subfigure}
    \caption{Starting from the blue line, we rewrite the chord diagram (left) by opening and closing chords at each node (right). We noted the state in the chord Hilbert space after each step for this particular diagram. The total weight of the diagram is $q^2$.}
    \label{fig:cutting chord diagram}
\end{figure}

The number of chord diagrams increases rapidly with $k$, so we need an efficient way of evaluating the sum over them. We do so by building all the chord diagrams iteratively with a \emph{transfer matrix} \cite{Berkooz:2018qkz} in the following way. First, choose some point on the circle of \autoref{fig:chord diagram} between the $k$ nodes, and declare that at this point there are no open chords. Next move clockwise around the circle and at each node either open a new chord or close an already open one. After $k$ steps we must return to the same point, and therefore must again have no open chords. All chord diagrams are uniquely generated by this process. An illustration of the process for a single diagram is given in \autoref{fig:cutting chord diagram}.

Suppose we have $n$ open chords as we go along the diagram. If a chord opens, then we put it at the bottom of the stack of open chords. We have no new intersections and have $n+1$ chords for the next step\footnote{There are other conventions for when intersections happen, resulting in conjugate transfer matrices.}. If a chord closes, it could be either the bottom one, adding no intersections, or the one above it, which would contribute a factor of $q$, or all the way to the top one, adding a factor of $q^{n-1}$. So summing over all possible diagrams, if a chord closes, we multiply by $1 + q + \cdots + q^{n-1} = \frac{1-q^n}{1-q}$, and reduce the number of chords by 1.
\begin{equation}
    \label{eq:transfer matrix fig}
    \tikzset{every picture/.style={line width=0.75pt}} 
    \vcenter{\hbox{\begin{tikzpicture}[x=0.75pt,y=0.75pt,yscale=-1,xscale=0.9]
        \path (0,50);
        \draw    (8.5,65) -- (85.5,65) ;
        \draw [line width=1.5]    (12.5,51) -- (48,51) ;
        \draw [line width=1.5]    (12.5,46) -- (48,46) ;
        \draw [line width=1.5]    (12.5,26) -- (48,26) ;
        \draw (30,69) node [anchor=north west][inner sep=0.75pt]   [align=left] {$\displaystyle |n\rangle $};
        \draw (27.3,33) node  [font=\normalsize] [align=left] {$\displaystyle \vdots $};
    \end{tikzpicture}}}
    \; \xrightarrow{\quad\text{\large $T$} \quad}\;
    \vcenter{\hbox{\begin{tikzpicture}[x=0.75pt,y=0.75pt,yscale=-1,xscale=0.9]
        \path (0,50);
        \draw [color={rgb, 255:red, 0; green, 0; blue, 0 }  ,draw opacity=1 ][line width=1.5]    (46,65.5) .. controls (46.25,60.25) and (49.75,56.5) .. (57.5,56) .. controls (65.25,56) and (77.25,56) .. (84,56) ;
        \draw [line width=1.5]    (46,51) -- (83.5,51) ;
        \draw [line width=1.5]    (46,46) -- (83.5,46) ;
        \draw [line width=1.5]    (47,26) -- (83.5,26) ;
        \draw    (8.5,66) -- (83.5,66) ;
        \draw [line width=1.5]    (10.5,51) -- (46,51) ;
        \draw [line width=1.5]    (10.5,46) -- (46,46) ;
        \draw [line width=1.5]    (10.5,26) -- (47,26) ;
        \draw (15,70) node [anchor=north west][inner sep=0.75pt]   [align=left] {$\displaystyle |n+1\rangle $};
        \draw (41.3,32.35) node  [font=\normalsize] [align=left] {$\displaystyle \vdots $};
    \end{tikzpicture}}}
    \; +
    \vcenter{\hbox{\begin{tikzpicture}[x=0.75pt,y=0.75pt,yscale=-1,xscale=0.9]
        \path (0,50);
        \draw [color={rgb, 255:red, 0; green, 0; blue, 0 }  ,draw opacity=1 ][line width=1.5]    (10.5,51) .. controls (18.25,51) and (28.5,51) .. (34.5,51) .. controls (40.5,51) and (47,52) .. (47,66) ;
        \draw [line width=1.5]    (46,46) -- (83.5,46) ;
        \draw    (8.5,66) -- (84.5,66) ;
        \draw [line width=1.5]    (10.5,46) -- (46,46) ;
        \draw [line width=1.5]    (47,26) -- (84.5,26) ;
        \draw [line width=1.5]    (10.5,26) -- (47,26) ;
        \draw (7.5,70) node [anchor=north west][inner sep=0.75pt]   [align=left] {$\displaystyle 1\cdot |n-1\rangle $};
        \draw (41.3,32.35) node  [font=\normalsize] [align=left] {$\displaystyle \vdots $};
    \end{tikzpicture}}}
    \; +
    \vcenter{\hbox{\begin{tikzpicture}[x=0.75pt,y=0.75pt,yscale=-1,xscale=0.9]
        \path (0,50);
        \draw [color={rgb, 255:red, 0; green, 0; blue, 0 }  ,draw opacity=1 ][line width=1.5]    (10.5,46) .. controls (20.5,46) and (33.5,46) .. (35.5,46) .. controls (47.5,46) and (46,47.5) .. (46,66) ;
        \draw [line width=1.5]    (46,51.5) -- (83.5,51) ;
        \draw    (8.5,66) -- (83.5,66) ;
        \draw [line width=1.5]    (10.5,52) -- (46,51.5) ;
        \draw [line width=1.5]    (44,26) -- (81.5,26) ;
        \draw [line width=1.5]    (10.5,26) -- (44,26) ;
        \draw (10.5,70) node [anchor=north west][inner sep=0.75pt]   [align=left] {$\displaystyle q\cdot |n-1\rangle $};
        \draw (41.3,31.35) node  [font=\normalsize] [align=left] {$\displaystyle \vdots $};
    \end{tikzpicture}}}
    \; + \;\cdots\; +
    \vcenter{\hbox{\begin{tikzpicture}[x=0.75pt,y=0.75pt,yscale=-1,xscale=0.9]
        \path (0,50);
        \draw [color={rgb, 255:red, 0; green, 0; blue, 0 }  ,draw opacity=1 ][line width=1.5]    (10.5,26) .. controls (21.5,26) and (29.5,26) .. (34.5,26) .. controls (47.5,26) and (47,48) .. (47,66) ;
        \draw [line width=1.5]    (46,51) -- (83.5,51) ;
        \draw [line width=1.5]    (46,46) -- (83.5,46) ;
        \draw    (8.5,66) -- (84.5,66) ;
        \draw [line width=1.5]    (10.5,51) -- (47,51) ;
        \draw [line width=1.5]    (10.5,46) -- (47,46) ;
        \draw (0,71) node [anchor=north west][inner sep=0.75pt]   [align=left] {$\displaystyle q^{n-1} \cdot |n-1\rangle $};
        \draw (24.3,32.35) node  [font=\normalsize] [align=left] {$\displaystyle \vdots $};
    \end{tikzpicture}}}
\end{equation}

This process is an evolution in the \emph{chord Hilbert space} 
\begin{equation} 
\label{eq:chord Hilbert space}
\cH = \left\{ \ket{n} \big| n \in \bZ_+ \right\}\,,
\end{equation}
which describes how many open chords there are at any given step. We start from "no chords", $\ket{0}$, and end with $\bra{0}$. Defining
\begin{equation}
    \label{eq:DefChordCreation}
\begin{aligned}
    \alpha \ket{n} &= \ket{n-1} \,, &\alpha^+ \ket{n} &= \ket{n+1} \,,  &\hat n \ket{n} &= n \ket{n} \,, \\
    a &= \alpha \frac{1 - q^{\hat n}}{1-q} \, ,  &a^\dagger &= \alpha^+  \,,  &T &= a^\dagger + a \,.
\end{aligned}
\end{equation}
the moments and partition function are
\begin{equation}\label{eq:TasSumAADeg}
    m_{k} = \left\langle \Tr(H^{k})\right\rangle_J = \langle 0 | T^{k} | 0 \rangle \,, \qquad   Z = \langle 0 | e^{-\beta T} | 0 \rangle \,.
\end{equation}
$T$ is called the \emph{transfer matrix}, and we can think about it as the Hamiltonian acting on the chord Hilbert space. Its spectrum and eigenstates are known, allowing us to compute this expression, as we will do shortly.

More conceptually, by the same arguments any quantum mechanical time evolution $e^{-iHt}$ inside an ensemble averaged correlation function can be replaced by $e^{-iTt}$ acting on the chord Hilbert space. So we have derived a duality explicitly -- computations in the microscopic Hilbert space are equal to those in the chord Hilbert space. This is just the AdS/CFT duality in this case, as we will see shortly that $T$ is the gravitational Hamiltonian.

This also is the first appearance of non-commutative geometry as the operators $a$, $a^\dagger$ are the Arik-Coon $q$-oscillators which satisfy
\begin{equation}\label{eq:ArikCoon}
[a, a^\dagger]_q \equiv a a^\dagger - q a^\dagger a = 1 \,, \qquad [{\hat n},a]=-a,\qquad [{\hat n},a^\dagger]=a^\dagger \,.
\end{equation}
We also choose an inner product on $\cH$ such that $a$ and $a^\dagger$ are Hermitian conjugates,
\begin{equation}
    \label{eq:inner_product_no_matter}
    \langle m | n \rangle = \delta_{mn} [n]_q! \,,
\end{equation}
where $[n]_q! = \frac{(q;q)_n}{(1-q)^n}$ is the $q$-factorial defined in \eqref{eq:q factorial def}, and $(q;q)_n = \prod_{k=1}^{n}(1-q^k)$ the $q$-Pochhammer symbol.

\paragraph{The thermofield double}
Even though we have a Hilbert space, we are not computing traces over it but rather transition amplitudes. A more careful interpretation, due to Lin \cite{Lin:2022rbf}, is that the chord Hilbert space describes the dynamics on the doubled Hilbert space of the system\footnote{See also \cite{Okuyama:2024yya} in the context of DS-SYK.} -- a tensor product of two copies of its Hilbert space, named the left- and the right-copy. We briefly introduce the construction, and return to it later in Section~\ref{sec:complexity}.

\begin{figure}[t]
    \centering
    \begin{subfigure}[t]{0.36\textwidth}
        \centering
        \includegraphics[width=1.0\textwidth]{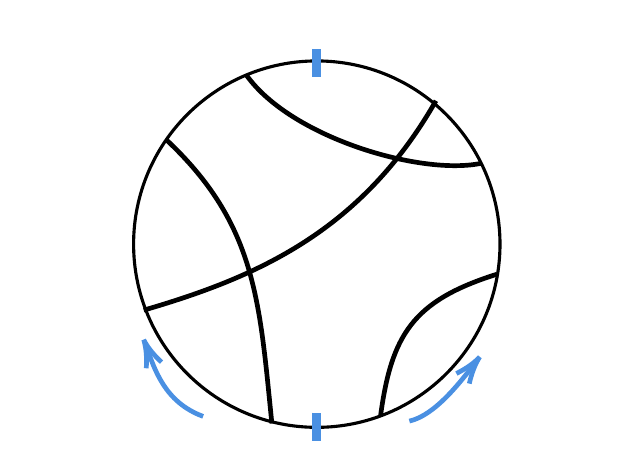}
        \subcaption{The cuts to the diagram.}
        \label{fig:chord diagram 2 cuts}
    \end{subfigure}
    \quad
    \begin{subfigure}[t]{0.6\textwidth}
        \centering
        \includegraphics[width=0.6\textwidth]{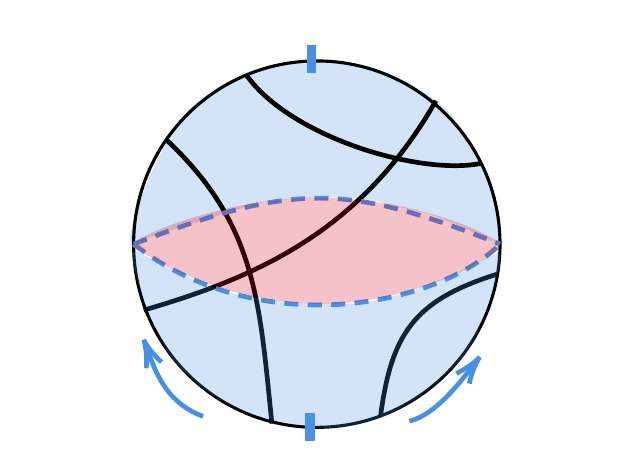}
        \subcaption{The contribution of a single diagram to the bra- and ket-states (blue) and their inner-product (red).}
        \label{fig:bra and ket in chord diagram}
    \end{subfigure}
    \caption{Chord diagrams as transition amplitude in the two-sided Hilbert space.}
    \label{fig:two_sided_cuts}
\end{figure}

Consider the state $\ket{0} = \sum_{k=1}^\infty \ket{k}_L \ket{k}_R$ on the doubled space, where $\{\ket{k}\}_{L,R}$ are two copies of a basis of the single space. Given two copies of the one-sided Hamiltonians $H_{L,R}$, there is another useful state, the thermofield double state $\ket{\text{TFD}} = e^{-\frac{\beta}{4}(H_L +H_R)} \ket{0}$. These can be used to express the partition function of the theory as an amplitude, and in DS-SYK once we ensemble average over the couplings it is
\begin{equation}
    Z = \big\langle \braket{0 | e^{-\frac{\beta}{2} \left(H_L + H_R\right)} | 0} \big\rangle_J = \big\langle \braket{\text{TFD} | \text{TFD}} \big\rangle_J\,,
\end{equation}
where the couplings of both Hamiltonians are drawn from the same distribution. This again results in a sum over chord diagrams, and has the pictorial representation of cutting them open in two places, \autoref{fig:chord diagram 2 cuts}. 
As before, all chord diagrams are generated by a transfer matrix \cite{Lin:2022rbf}. At each step, the state of the system is given by the number of open chords, $\ket{n}$. Then, a chord can either open from the left boundary or close into it, and similarly from the right. As a chord closes, it crosses all the chords between itself and the appropriate boundary. Our transfer matrix now has two parts, $T_L$ and $T_R$, in terms of which the partition function and TFD are
\begin{equation}
    Z = \braket{0| e^{-\frac{\beta}{2}(T_L + T_R)} | 0} \,, \qquad \ket{\rm TFD} = e^{-\frac{\beta}{4}(T_L + T_R)}\ket{0} \,.
\end{equation}

The TFD state can be thought of as a (weigthed) sum over all ``bottom halves'' of a chord diagram, without accounting for the intersections between the chords that leave them. In the amplitude $\braket{\text{TFD}|\text{TFD}}$ the inner product \eqref{eq:inner_product_no_matter} accounts for these intersections given two such halves\footnote{Say there are $n$ chords coming out of each half, and look at them from left to right. When the first chord on the bottom connects to the first chord on the top, it does not intersect any of the other chords. When it connects to the second top chord, it intersects exactly one chord and gives a factor of $q$. Summing over the ways to connect it gives a factor of $[n]_q$. The second bottom chord then gives an additional factor of $[n-1]_q$, and so it goes until we get an overall factor of $[n]_q!$.}, \autoref{fig:bra and ket in chord diagram}, 
thus summing over all diagrams.

\paragraph{Diagonalizing the transfer matrix} 
The transfer matrix $T$ is Hermitian under the inner product \eqref{eq:inner_product_no_matter}, and admits a complete set of eigenstates with real eigenvalues. The spectrum is continuous and bounded, such that the eigenvalues are in $[-\frac{2}{\sqrt{1-q}},\frac{2}{\sqrt{1-q}}]$, and it can be parameterized by an angle $\theta \in [0,\pi]$. We denote by $\ket{\theta}$ the eigenvector and $\psi_n(\theta)$ is its $n^{\text{th}}$ component in the chord basis, such that
\begin{equation}
    \label{eq:energy eigenstates}
    T \ket{\theta} = \frac{2\cos(\theta)}{\sqrt{1-q}} \ket{\theta} \,,\qquad \psi_n(\theta) = \sqrt{\rho(\theta)} \frac{(1-q)^{n/2}}{(q;q)_n}H_n\left(\cos(\theta)|q\right) \,,
\end{equation}
where $(q;q)_n$ denotes the $q$-Pochhammer symbol, $H_n(x|q)$ the continuous $q$-Hermite polynomials, and $\rho(\theta) = \frac{(q,e^{\pm 2i\theta};q)_\infty}{2\pi}$, which we will soon interpret as the density of states, is a function that ensures the completeness relation
\begin{equation}
    \int_0^{\pi} d\theta \,  \langle n| \theta \rangle  \langle \theta | m \rangle = [n]_q! \delta_{nm} \quad \implies \quad \int_0^\pi d\theta |\theta\rangle \langle \theta| = \1 \,.
\end{equation}
The list of relevant mathematical definitions is in Appendix \ref{app:MathDef}.

\begin{figure}
    \centering
    \begin{subfigure}{0.25\textwidth}
        \centering
        \includegraphics[width=1\textwidth]{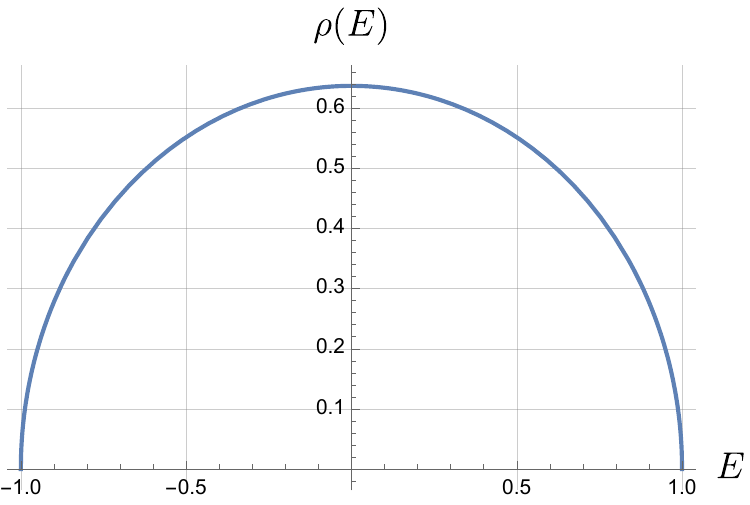}
    \end{subfigure}
    \qquad
    \begin{subfigure}{0.25\textwidth}
        \centering
        \includegraphics[width=1\textwidth]{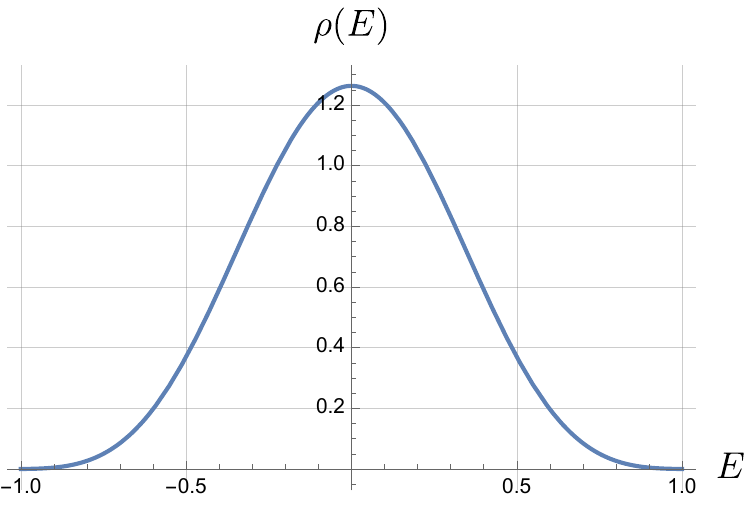}
    \end{subfigure}
    \qquad
    \begin{subfigure}{0.25\textwidth}
        \centering
        \includegraphics[width=1\textwidth]{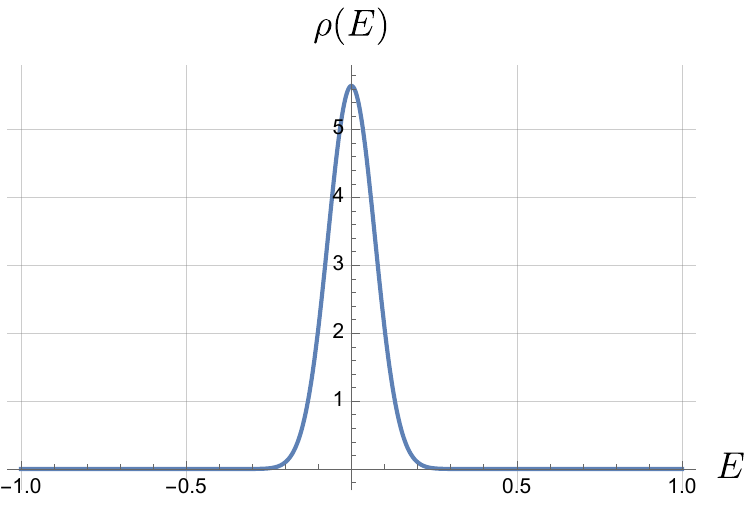}
    \end{subfigure}

    \begin{subfigure}{0.25\textwidth}
        \centering
        \includegraphics[width=1\textwidth]{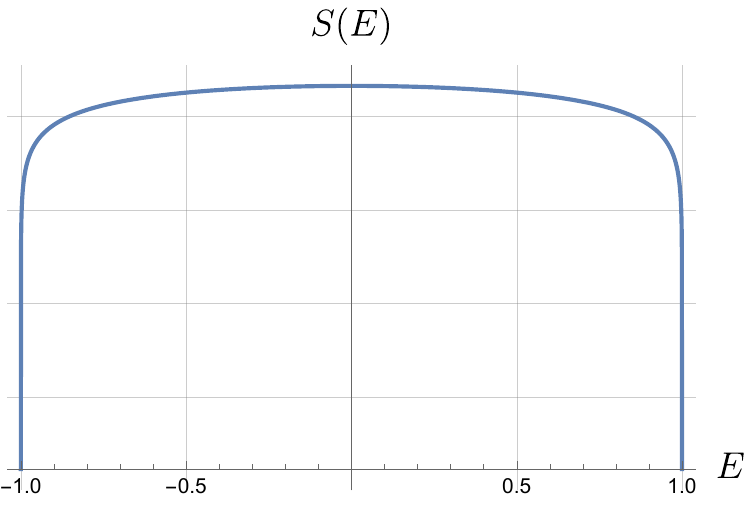}
        \subcaption{$q = 0.01$}
    \end{subfigure}
    \qquad
    \begin{subfigure}{0.25\textwidth}
        \centering
        \includegraphics[width=1\textwidth]{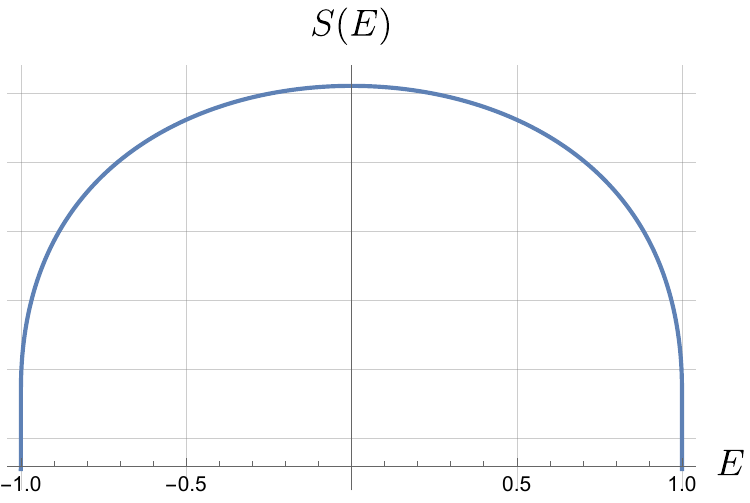}
        \subcaption{$q = 0.8$}
    \end{subfigure}
    \qquad
    \begin{subfigure}{0.25\textwidth}
        \centering
        \includegraphics[width=1\textwidth]{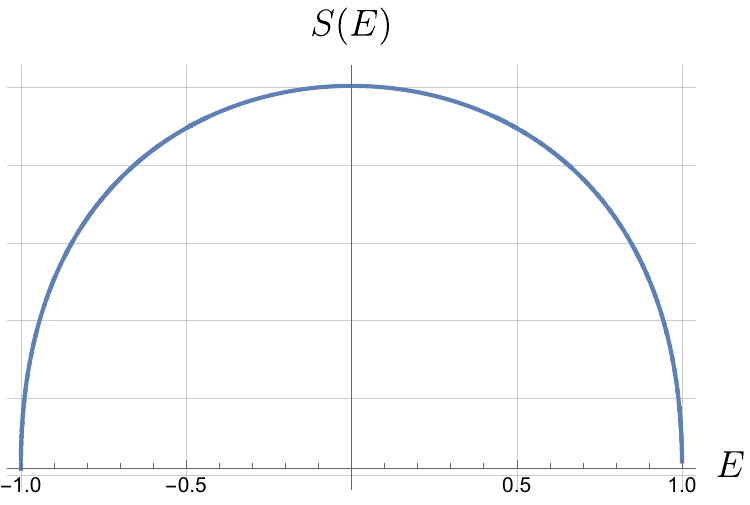}
        \subcaption{$q = 0.99$}
    \end{subfigure}
    
    \caption{The density of states (top) in units of $2^{N/2}$ and the entropy (bottom) for different values of $q$. The energies are in units of $\frac{2\bJ}{\sqrt{\lambda(1-q)}}$. As discussed in Footnote~\ref{foo: trace norm}, we subtract a constant $\frac{N}{2}\log(2)$ from the entropy.}
    \label{fig:density_of_states}
\end{figure}

\paragraph{The spectrum}
The partition function can now be phrased as a one dimensional integral by inserting complete sets of eigenstates of $T$ into \eqref{eq:TasSumAADeg},
\begin{equation}
    \label{eq:exact partition function}
    Z = \int_0^{\pi} d\theta \int_0^{\pi} d\theta' \, \langle 0|\theta\rangle \langle \theta | e^{-\beta T} | \theta'\rangle \langle \theta' | 0\rangle =  \int_0^{\pi} \frac{d\theta}{2\pi} (q,e^{\pm2i\theta};q)_\infty e^{-\frac{2\beta\bJ\cos(\theta)}{\sqrt{\lambda(1-q)}}} \,,
\end{equation}
where we reintroduced the dimensionful coupling $\cJ=\frac{\bJ}{\sqrt{\lambda}}$ that was set to $1$ earlier. The spectrum is bounded and parameterized by $\theta$, and the density of states is \cite{Garcia-Garcia:2017pzl,Cotler:2016fpe,Berkooz:2018jqr}
\begin{equation}
\begin{gathered}
    \label{eq:density of states poc}
    \rho(\theta)d\theta = \frac{1}{2\pi}(q,e^{\pm 2i\theta}; q)_\infty d\theta\,, \qquad E(\theta) = \frac{2\bJ\cos(\theta)}{\sqrt{\lambda(1-q)}} \,, \qquad \theta \in [0,\pi]\,.\\
    \text{or} \quad \rho(E)dE =  \frac{\sqrt{\lambda(1-q)}}{2\pi\bJ} \frac{1}{q^{1/8}} \theta_1\left(\frac{\theta(E)}{\pi} \bigg| \frac{\lambda i}{2\pi}\right)dE\,,
\end{gathered}
\end{equation}
where we used \eqref{eq:Jacobi_theta_series}. \autoref{fig:density_of_states} shows the shape of $\rho(E)$ for different values of $q$. After using the modular transform \eqref{eq:Jacobi_theta_modular} followed by the expansion \eqref{eq:Jacobi_theta_series} we find the form \cite{Verlinde:2024znh}
\begin{equation}
    \label{eq:density_of_states_modular}
    \rho(\theta) = \frac{\sin\left(\theta\right)}{\pi q^{1/8}}\sqrt{\frac{2\pi}{\lambda}}\sum_{k=-\infty}^{\infty}\left(-1\right)^{k}e^{-\frac{2}{\lambda}\left(\theta-\pi\left(k+\frac{1}{2}\right)\right)^{2}} \,,
\end{equation}
which is valid for any $q$, and almost suggests a full semi-classical interpretation.

It should be stressed that in the double scaling limit the size of the Hilbert space is infinite, and so is the (physical) density of states. However, since our normalization of the trace is $\Tr(\1) = 1$, the density of states $\rho(E) dE$ we compute describes what fraction of the Hilbert space has energy within a window of size $dE$ around $E$, and that remains finite even in this limit.

\subsection{Adding matter and two point functions}

To probe the quantum state of the system further we would like to come up with a suitable set of probes \cite{Berkooz:2018qkz}. We have already considered one operator -- the Hamiltonian \eqref{eq:SYK Hamiltonian} -- which is drawn from a specific statistical ensemble of operators. Next we would like to define the notion of an appropriate probe operator. The correct, or at least convenient, choice is to draw probe operators from similar ensembles modified to encode the fact that they may have different dimensions (and other quantum numbers, like charges, if present in the model).


In the context of the $AdS/CFT$ duality, and in particular if we think of a dual 2D black holes as arising from a near horizon S-wave compactification of a higher dimensional black hole, then these probes should be the analogues of the "single trace" operators. 
In this context, a simple ``single trace'' operator in the UV theory, such as the Hamiltonian, flows to a very complicated operator \eqref{eq:SYK Hamiltonian} in terms of the effective description of the IR theory, which we take to be the fermions of the SYK model. Similarly, other simple ``single trace'' operators in the UV will look complicated in the IR, but their overall renormalized appearance need not be that different from that of the Hamiltonian.

The easiest way to achieve this is to take operators similar to \eqref{eq:SYK Hamiltonian}, except with a different length $\tilde p$ instead of $p$ and with a new set of independent Gaussian couplings, 
\begin{equation}
    \label{eq:RandOp}
    M_{\Delta} = i^{\tilde p/2} \sum_{i_1 < i_2 <\cdots < i_{\tilde p} = 1}^{N} {\tilde J}_{i_1 \cdots i_{\tilde p}} \psi_{i_1} \cdots \psi_{i_{\tilde p}} \equiv i^{\tilde p/2} \sum_{I'} \tilde J_{I'} \Psi_{I'}  \,.
\end{equation}
We denoted this operator by $\Delta=\tilde p/p$ since in the JT gravity limit of Section~\ref{sec:triple scaling} it corresponds to an operator with conformal dimension $\Delta$. However, we can evaluate its exact (unnormalized) thermal two point function before taking such a limit.
We want to compute 
\begin{equation}
    G_\Delta(\beta_1,\beta_2) = \left\langle \Tr\left[e^{-\beta_1 H} M e^{-\beta_2 H} M\right] \right\rangle_{J,\tilde J} = \big\langle \Tr\big[e^{-(\beta_1 + \beta_2) H}  M(\beta_2) M(0)\big] \big\rangle_{J,\tilde J}\,,
\end{equation}
or, after expanding into moments, $\left\langle \Tr( H^{k_1} M  H^{k_2} M) \right\rangle_{J,\tilde J}$, where the ensemble average is over both $J$ and $\tilde J$. The chord diagram associated with this observable is one in which we pair the $H$'s as before, but there is one distinct chord which connects the two $M$'s -- we will call these $H$-chords and $M$-chord. Again as before, the intersection of two $H$-chords comes with a weight $q$ but the intersection of an $H$-chord with the $M$-chord comes with a new weight\footnote{The combinatorics are a simple generalization of Footnote~\ref{foo: Poisson dist} to the case where the two index sets are of different lengths.} $\tilde q$
\begin{equation}
q=e^{-{2p^2 \over N}},\qquad {\tilde q}=e^{-{2p\tilde p\over N}} = q^{\Delta} \,.
\end{equation}

\begin{figure}[t!]
    \centering
    \begin{subfigure}{0.5\textwidth}
        \centering
        \includegraphics[width=0.5\textwidth]{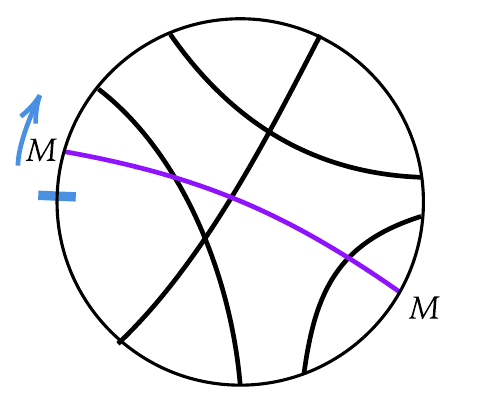}
    \end{subfigure}
    \quad
    \begin{subfigure}{0.45\textwidth}
        \centering
        \includegraphics[width=\textwidth]{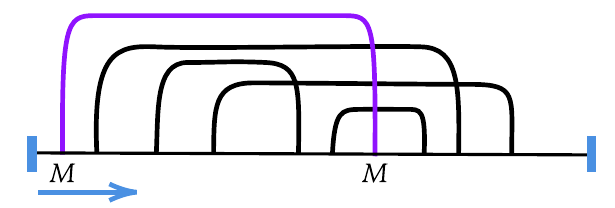}
    \end{subfigure}
    \caption{Marked chord diagram for a two point function.}
    \label{fig:Chord diagram for Two point function}
\end{figure}

A variant of the transfer matrix method works here just as well. We cut the circular diagram such that one $M$ appears just to the right of the cut, and the other $M$ somewhere in the interior, as in \autoref{fig:Chord diagram for Two point function}. The transfer matrix takes care of the ``bubbling" of $H$-chords and it is the same as before. We start with the leftmost insertion of $H$ and propagate the system through the first $k_2$ steps, i.e. the $k_2$ $H$'s between the two $M$ operators, which gives the same contribution as before. However, when we hop over the second insertion of $M$, the open $H$-chords cross the $M$-chord picking up an additional factor of ${\tilde q}^{\,\text{no.\ of\ lines}}$. Finally, we propagate the chords for the remaining $k_1$ steps. The expression for 2-pt function is therefore
\begin{equation}
\label{eq:2 pt function transition amp}
	\left\langle \Tr\bigl( e^{-\beta_1 H}MH^{-\beta_2 H}M\bigr) \right\rangle_{J,\tilde J} = \langle 0 | e^{-\beta_1 T} \tilde q^{\,\hat n} e^{-\beta_2 T} | 0 \rangle \,.
\end{equation}
After inserting complete sets of energy states \eqref{eq:energy eigenstates} and using \eqref{eq:q_hermite_n_orthogonality} one finds \cite{Berkooz:2018qkz} 
\begin{equation}\label{2PTTheta}
	G_\Delta(\beta_1,\beta_2)
	 =
 \int_0^\pi d\theta_1 \int_0^\pi d\theta_2 \, \rho(\theta_1) \rho(\theta_2) \, e^{-\beta_1 E(\theta_1) - \beta_2 E(\theta_2)}  \frac{({\tilde q}^2;q)_\infty}{(\tilde q e^{i(\pm \theta_1 \pm \theta_2)};q)_\infty} \,.
\end{equation}

\subsection{The $p$-spin model}\label{sec:Pspin}

Whenever the dynamics of a system is described via chord diagrams, then the transfer matrix and the subsequent machinery can be applied to it. This is expected to be the case for many double scaled $p$-local systems described in Section~\ref{sec:Generalizations} (and even beyond, see Section~\ref{sec:parisi}), so it should be viewed more as a broad universality class, rather than a trick for a specific model. In fact, equation \eqref{eq:MmntsChrds} was actually written first for the $p$-spin model \cite{Erdos}, and only then adapted to the SYK model \cite{Cotler:2016fpe}. The entire machinery of the transfer matrix was applied to the $p$-spin model first in \cite{Berkooz:2018qkz}.

The $p$-spin model is the following. Consider $N$ qubits, and Pauli matrices $\sigma^{(a_i)}_i$ with $a_i=1,2,3$ acting on the $i$-th qubit, $i=1,\dots,N$. Denote by $e=(i_1,\dots,i_p)$ some subset of the qubits, and by $A=(a_1,\dots,a_p)$ the choice of each Pauli matrix acts on each of them. Denoting the pair $(e,A)$ by $I$, and a tensor product of the matrices by 
\begin{equation}\label{RandHam}
	\sigma_I=\sigma_{(e,A)}=\sigma_{i_1}^{(a_1)}\sigma_{i_2}^{(a_2)}\cdots\sigma_{i_p}^{(a_p)} \,,
\end{equation}
we define the random Hamiltonian with Gaussian random couplings $J_I$ to be
\begin{equation}\label{eq:SpinHamilt}
	H_{p\text{-spin}} = 3^{-p/2}{n\choose p}^{-1/2}\sum_{I} J_I \sigma_I\,.
\end{equation}
The relevant parameter controlling the double scaling limit $N\rightarrow\infty$, $\lambda$ fixed is \cite{Erdos}
\begin{equation}\label{ParamDef}
	q=e^{-\lambda} \quad \text{with} \quad \lambda= \frac{4}{3}\ \frac{p^2}{N} \ ,
\end{equation}

The computation of the moments, 
 $\sum_{I_1,\cdots,I_k} \bigl\langle J_{I_1}\cdots J_{I_k} \bigr\rangle_J \Tr\bigl( \sigma_{I_1}\cdots\sigma_{I_k} \bigr)$,
proceeds very similarly to the SYK model. First we carry out the ensemble average by identifying pairs of $I$'s, which brings us to the chord picture. Next, however, we do not ``pass one interaction term across another", but recall that in the double scaling limit each qubit appears in at most two chords.
The number of times the same qubit is shared by two index sets $I_1$ and $I_2$ is Poisson distributed with expectation value ${\frac{3\lambda}{4}}$.

\sloppy Now we carry out the trace on each qubit Hilbert space, with four cases to consider. The first is when a qubit does not appear in any operator. It then contributes 1 to the normalized trace. Second, a qubit that only appears in one chord gives a weight\footnote{The factor of $1/2$ is because we normalize $\tr_i(\1)=1$.} of $3^{-1} \sum_a {1\over 2} \tr_i\left(\sigma^{(a)}\sigma^{(a)}\right)=1$. Next, a qubit that appears in two non intersecting chord gives $3^{-2}\sum_{a,b=1}^3 {1\over 2} \tr_i\bigl( \sigma^{(a)}\sigma^{(a)}\sigma^{(b)}\sigma^{(b)} \bigr)~=~1$. Finally, a qubit that appears in two intersecting chords gives $3^{-2}\sum_{a,b=1}^3 {1\over 2} \tr_i\bigl( \sigma^{(a)}\sigma^{(b)}\sigma^{(a)}\sigma^{(b)} \bigr)= -{1\over 3}\,.$
\sloppy Recalling the Poisson nature of $|I_1\cap I_2|$, each pairwise intersection gives a factor of $\langle (-{1 \over 3})^{|I_1\cap I_2|}\rangle_{\text{Poiss},{3 \lambda\over 4}} =q$, and we are back in equation \eqref{eq:MmntsChrds} for this model as well.

\section{Semiclassics (bulk I)}\label{HologA} 
\label{sec:semi-classics}

\subsection{The triple scaling limit and the Schwarzian theory}
\label{sec:triple scaling}
At low energies the SYK model with finite $p$ is described by the Schwarzian theory (and JT gravity) \cite{kitaev2015simple, Maldacena:2016hyu}. This limit also arises in our case by taking the \emph{triple scaling limit} \cite{Cotler:2016fpe}, which is when we take $\lambda \ll 1$ while concentrating on energies near the edge of the spectrum.

More precisely, suppose we write $\theta = \pi - \lambda \sqrt{\varepsilon}$, such that $E - E_{\text{min}} = \lambda \bJ  \varepsilon$ when $\lambda \to 0$. After the modular transformation \eqref{eq:Jacobi_theta_modular} and the series expansion \eqref{eq:Jacobi_theta_series} the density of states, \eqref{eq:density of states poc}, reproduces that of the Schwarzian theory up to an overall factor,
\begin{equation}
    \rho(\varepsilon) = \sqrt{\frac{2\lambda^3}{\pi}}e^{-\frac{\pi^2}{2\lambda}} \sinh\left(2\pi \sqrt{\varepsilon}\right) \,.
\end{equation}
As we will discuss below, the Schwarzian behavior can also be obtained in the canonical ensemble, when $\lambda \to 0$ the temperature is low, $\beta\bJ \gg 1$. The two-point function for an operator of length $\tilde p$ can also be computed in this limit, and it is proportional\footnote{At even lower temepratures, where $\lambda\beta\bJ \gg 1$, the propagator has a slightly different form. This is the very-low temperature regime discussed in
\cite{Berkooz:2018jqr}.} to ${Z^{-1} G_\Delta(\beta-\tau,\tau) = \sin^{-2\Delta}(2\pi \tau/\beta)}$, i.e. the two point function of a conformal operator of dimension $\Delta = \tilde p / p$.

We can do more than just reproduce the density of states, and match with a (form of) the Schwarzian directly \cite{Berkooz:2018qkz}. First conjugate the transfer matrix to the form
\begin{equation}
T = {1\over \sqrt{1-q}} \begin{pmatrix} 
0& 1 &0 & 0 & \cdots \\
1 & 0 & 1-q & 0 & \ddots \\
0 & 1 & 0 & 1-q^2 & \ddots\\
\vdots & \ddots & \ddots& \ddots & \ddots
\end{pmatrix} \,.
\end{equation}
When we go to large values of $n$ (way down the diagonal) and replace it by a continuous variable, the 1's above and below the diagonal go over the 2nd derivative, and the $q^n$ becomes a potential. The transfer matrix essentially describes dynamics on a discrete one sided lattice.

Using the notation of \cite{Lin:2022rbf,Blommaert:2024ydx} we can define the new operators, $\ell$, and its (formal) conjugate $\pi_\ell$, and write the transfer matrix as\footnote{Formally, this is related to \eqref{eq:DefChordCreation} by $\alpha = -\sqrt{1-q}\,e^{-i\lambda \pi_\ell}$, $\alpha^+ = -\frac{1}{\sqrt{1-q}}\,e^{i\lambda \pi_\ell}$, but $\pi_{\ell}$ is not entirely well defined at any finite $\lambda$, since $\ell$ is discrete.} 
\begin{equation} \label{eq:q-Liouville Hamiltonian}
    \ell = \lambda \hat n \,, \qquad [\pi_\ell, \ell] = i \,, \qquad   T = -\frac{\bJ}{\sqrt{\lambda(1-q)}} \left(e^{i\lambda \pi_\ell} + e^{-i\lambda \pi_\ell}\left(1-e^{-\ell}\right) \right) \,.
\end{equation}
Going to low energies/long wavelength/$\ell\to\infty$ on the lattice, we define $\tilde \ell = \ell + 2\log\lambda$ and keep it fixed when $\lambda\rightarrow 0$. The transfer matrix then becomes
\begin{equation}
    \label{eq:triple_scaling_Liouville_Ham}
    T = E_{\text{min}} + \frac{\lambda\bJ}{2} \left(\pi_{\tilde\ell}^2 + e^{-\tilde\ell} \right) \,, \qquad E_{\text{min}} = -\frac{2\bJ}{\lambda} \,.
\end{equation}
This is exactly the Liouville Hamiltonian which is a gauge fixed version of the Schwarzian quantum mechanics \cite{Bagrets:2016cdf, Bagrets:2017pwq}, and so the transfer matrix that captures the dynamics on this new emergent chord degrees of freedom becomes the gravitational Hamiltonian. In the thermofield double interpretation, the Hilbert space is that of JT-gravity on a strip with two boundaries \cite{Harlow:2018tqv}, where $\ell$ and $\tilde \ell$ play the role of the bare and renormalized lengths between the two boundaries, respectively.

Note that for finite $\lambda$ this length is quantized by $\lambda$. A theory which has a minimal length scale manifest throughout the computation
is entitled to be considered a full theory of \emph{Quantum Gravity}. This short distance is usually expected to be the Planck scale\footnote{The role of a string scale here is unclear since there is no obvious stringy description of this background.}. A little more precisely, the semiclassical two point function of an operator of dimension $\Delta$ is  
\begin{equation}
{\tilde q}^{\,\hat n} = e^{-\lambda \Delta \cdot \hat n } \sim e^{-m R_{AdS}\cdot \ell}
\end{equation}
so changing $n$ by 1 means that we change the distance by $\lambda R_{AdS}$, i.e. $\ell_{Planck}=\lambda R_{AdS}$. 

Recall also that the number of states is $2^{N/2}$ which is formally infinite in the double scaling limit, so the situation is similar to the limit of an infinitely large extremal black hole with $\ell_p$ kept fixed. The early time behavior of the partition function, two-point function, and OTOC (as we will see below) is controlled by $\lambda$, while the size of the Hilbert space is controlled by $N$. Since we can reinstate $N$ in all the expressions, we will discuss subleading effects that are sensitive to $N$ in Section~\ref{sec: wormholes}.

\subsection{Finite and fake temperature}
\label{sec:finite_temp}

We will now summarize some features of the model when $\lambda \to 0$ but the temperature $\beta\bJ$ is finite, such that we are sensitive to the entire spectrum. The two-point function and the Lyapunov exponent will behave almost holographically, the only difference being that the temperature dependence will be replaced by a ``fake'' temperature. The leading order of these quantities reproduces\footnote{At high temperatures $\beta\bJ \ll \lambda$, the $\beta\bJ$ expansion is just the moment expansion, allowing to compare the subleading orders in $\lambda$ for the double scaling limit and in $1/p$ for the large $p$ limit, and to verify that they differ \cite{Garcia-Garcia:2018fns, Jia:2018ccl, Berkooz:2018jqr}. } the large $p$ limit of SYK \cite{Maldacena:2016hyu}. At low temperatures, the fake temperature converges to the real one, and the triple scaling limit is reproduced.

\paragraph{The partition function}
Consider the density of states in the form \eqref{eq:density_of_states_modular}, and focus on the dominant\footnote{For other $k$'s the story is similar when defining $\theta-\frac{\pi\left(2k+1\right)}{2} = \left(-1\right)^{k}\frac{\pi v_{k}}{2}$ with $\theta \in [0,\pi]$.} term, $k=0$. 
At the saddle point, the average energy can be parameterized by the variable $v$, and the equations become
\begin{equation}
    \label{eq:def v and saddle point}
    \frac{\pi v}{\cos\left(\frac{\pi v}{2}\right)} = \beta\bJ \,,\qquad \theta=\frac{\pi}{2}+\frac{\pi v}{2}\,, \qquad v\in[-1,1] \,,
\end{equation}
resulting in \cite{Goel:2023svz,Mukhametzhanov:2023tcg,Okuyama:2023bch} 
\begin{equation}
    Z(\beta) = \frac{\cos\left(\frac{\pi v}{2}\right)}{\sqrt{1+\frac{\pi v}{2}\tan\left(\frac{\pi v}{2}\right)}}\exp\left[-\frac{1}{\lambda}\left(\frac{\pi^{2}v^{2}}{2}-2\pi v\tan\left(\frac{\pi v}{2}\right)\right)+\frac{1}{2}\pi v\tan\left(\frac{\pi v}{2}\right)\right]\,.
\end{equation}
At low temperatures $\beta\bJ\rightarrow\infty$ it agrees with the Schwarzian result\footnote{The first term in the exponent comes from the ground state energy of the model, ${E_{\text{min}} = -\frac{2\bJ}{\sqrt{\lambda(1-q)}}}$. The $\beta$ independent terms are scheme dependent in the Schwarzian theory, but not in DSSYK.} \cite{Maldacena:2016hyu} for the action $C\int_0^\beta d\tau \left\{\tan\left(\frac{\pi f(\tau)}{\beta}\right),\tau\right\}$, where $\{\cdot,\cdot\}$ is the Schwarzian derivative and the coefficient is $C = \frac{1}{2\lambda\bJ}$,
\begin{equation}
\label{eq:low_temp_partition_func}
    Z\left(\beta\right) = \frac{\pi\sqrt{2}}{\left(\beta\bJ\right)^{3/2}}e^{\left(\frac{2}{\lambda} + \frac{1}{2}\right)\beta\bJ-\frac{\pi^{2}}{2\lambda}+\frac{\pi^{2}}{\beta\bJ\lambda}} \,.
\end{equation}

\paragraph{The two point function}
Again by saddle point methods, from \eqref{2PTTheta} follows the normalized two-point function of a random operator of finite dimension $\Delta = \frac{\tilde p}{p} =  \frac{\tilde\lambda}{\lambda}$ \cite{Goel:2023svz,Mukhametzhanov:2023tcg,Okuyama:2023bch},
\begin{equation}
    \label{eq:finite temperature 2 pt}
    \frac{G_\Delta(\beta-\tau,\tau)}{Z} = \left[\frac{\cos\left(\frac{\pi v}{2}\right)}{\cos\left[\pi v\left(\frac{1}{2} - \frac{\tau}{\beta}\right)\right]}\right]^{2\Delta} \,.
\end{equation}
For the first correction in $\lambda$ see \cite{Okuyama:2023bch}. At low temperatures $v \to 1$ the numerator vanishes, forcing us to renormalize our operators, which is related to the renormalization of the bulk length as we will see below. Up to this factor, \eqref{eq:finite temperature 2 pt} matches the Schwarzian theory \cite{Okuyama:2023bch}.

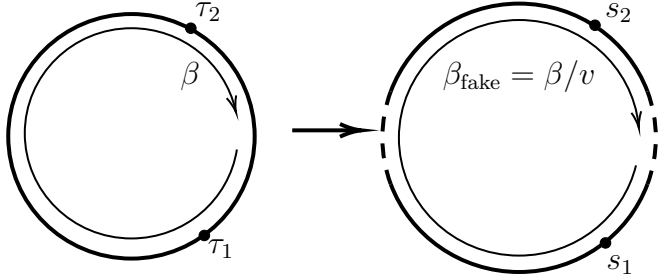
\begin{figure}[t]
    \centering
    \begin{tikzpicture}[x=0.75pt,y=0.75pt,yscale=-0.75,xscale=0.75]
        \draw  [draw opacity=0][line width=1.5]  (445.42,136.09) .. controls (432.63,170.48) and (399.21,195) .. (360,195) .. controls (320.79,195) and (287.37,170.48) .. (274.58,136.09) -- (360,105) -- cycle ; \draw  [line width=1.5]  (445.42,136.09) .. controls (432.63,170.48) and (399.21,195) .. (360,195) .. controls (320.79,195) and (287.37,170.48) .. (274.58,136.09) ;  
        \draw  [draw opacity=0][line width=1.5]  (274.58,73.91) .. controls (287.37,39.52) and (320.79,15) .. (360,15) .. controls (399.21,15) and (432.63,39.52) .. (445.42,73.91) -- (360,105) -- cycle ; \draw  [line width=1.5]  (274.58,73.91) .. controls (287.37,39.52) and (320.79,15) .. (360,15) .. controls (399.21,15) and (432.63,39.52) .. (445.42,73.91) ;  
        \draw  [line width=1.5]  (20,104) .. controls (20,58.71) and (56.71,22) .. (102,22) .. controls (147.29,22) and (184,58.71) .. (184,104) .. controls (184,149.29) and (147.29,186) .. (102,186) .. controls (56.71,186) and (20,149.29) .. (20,104) -- cycle ;
        \draw [line width=1.5]    (209,100) -- (252,100) ;
        \draw [shift={(255,100)}, rotate = 180] [color={rgb, 255:red, 0; green, 0; blue, 0 }  ][line width=1.5]    (14.21,-4.28) .. controls (9.04,-1.82) and (4.3,-0.39) .. (0,0) .. controls (4.3,0.39) and (9.04,1.82) .. (14.21,4.28)   ;
        \draw  [fill={rgb, 255:red, 0; green, 0; blue, 0 }  ,fill opacity=1 ] (421,175.5) .. controls (421,173.57) and (419.43,172) .. (417.5,172) .. controls (415.57,172) and (414,173.57) .. (414,175.5) .. controls (414,177.43) and (415.57,179) .. (417.5,179) .. controls (419.43,179) and (421,177.43) .. (421,175.5) -- cycle ;
        \draw  [draw opacity=0][dash pattern={on 5.63pt off 4.5pt}][line width=1.5]  (445.42,73.91) .. controls (449.03,83.6) and (451,94.07) .. (451,105) .. controls (451,115.93) and (449.03,126.4) .. (445.42,136.09) -- (360,105) -- cycle ; \draw  [dash pattern={on 5.63pt off 4.5pt}][line width=1.5]  (445.42,73.91) .. controls (449.03,83.6) and (451,94.07) .. (451,105) .. controls (451,115.93) and (449.03,126.4) .. (445.42,136.09) ;  
        \draw  [draw opacity=0][dash pattern={on 5.63pt off 4.5pt}][line width=1.5]  (274.58,136.09) .. controls (270.97,126.4) and (269,115.93) .. (269,105) .. controls (269,94.07) and (270.97,83.6) .. (274.58,73.91) -- (360,105) -- cycle ; \draw  [dash pattern={on 5.63pt off 4.5pt}][line width=1.5]  (274.58,136.09) .. controls (270.97,126.4) and (269,115.93) .. (269,105) .. controls (269,94.07) and (270.97,83.6) .. (274.58,73.91) ;  
        \draw  [fill={rgb, 255:red, 0; green, 0; blue, 0 }  ,fill opacity=1 ] (414,29.5) .. controls (414,27.57) and (412.43,26) .. (410.5,26) .. controls (408.57,26) and (407,27.57) .. (407,29.5) .. controls (407,31.43) and (408.57,33) .. (410.5,33) .. controls (412.43,33) and (414,31.43) .. (414,29.5) -- cycle ;
        \draw  [fill={rgb, 255:red, 0; green, 0; blue, 0 }  ,fill opacity=1 ] (154,170.5) .. controls (154,168.57) and (152.43,167) .. (150.5,167) .. controls (148.57,167) and (147,168.57) .. (147,170.5) .. controls (147,172.43) and (148.57,174) .. (150.5,174) .. controls (152.43,174) and (154,172.43) .. (154,170.5) -- cycle ;
        \draw  [fill={rgb, 255:red, 0; green, 0; blue, 0 }  ,fill opacity=1 ] (145,31.5) .. controls (145,29.57) and (143.43,28) .. (141.5,28) .. controls (139.57,28) and (138,29.57) .. (138,31.5) .. controls (138,33.43) and (139.57,35) .. (141.5,35) .. controls (143.43,35) and (145,33.43) .. (145,31.5) -- cycle ;
        \draw  [draw opacity=0][line width=0.75]  (437.79,123.5) .. controls (429.36,158.21) and (397.73,184) .. (360,184) .. controls (315.82,184) and (280,148.63) .. (280,105) .. controls (280,61.37) and (315.82,26) .. (360,26) .. controls (401.97,26) and (436.39,57.92) .. (439.73,98.51) -- (360,105) -- cycle ; \draw [line width=0.75]    (437.79,123.5) .. controls (429.36,158.21) and (397.73,184) .. (360,184) .. controls (315.82,184) and (280,148.63) .. (280,105) .. controls (280,61.37) and (315.82,26) .. (360,26) .. controls (401.34,26) and (435.36,56.97) .. (439.56,96.69) ; \draw [shift={(439.73,98.51)}, rotate = 260.85] [color={rgb, 255:red, 0; green, 0; blue, 0 }  ][line width=0.75]    (10.93,-3.29) .. controls (6.95,-1.4) and (3.31,-0.3) .. (0,0) .. controls (3.31,0.3) and (6.95,1.4) .. (10.93,3.29)   ; 
        \draw  [draw opacity=0][line width=0.75]  (172.17,112.84) .. controls (166.92,146.63) and (137.5,172.5) .. (102,172.5) .. controls (62.79,172.5) and (31,140.94) .. (31,102) .. controls (31,63.06) and (62.79,31.5) .. (102,31.5) .. controls (136.38,31.5) and (165.05,55.76) .. (171.6,87.98) -- (102,102) -- cycle ; \draw [line width=0.75]    (172.17,112.84) .. controls (166.92,146.63) and (137.5,172.5) .. (102,172.5) .. controls (62.79,172.5) and (31,140.94) .. (31,102) .. controls (31,63.06) and (62.79,31.5) .. (102,31.5) .. controls (135.69,31.5) and (163.9,54.8) .. (171.18,86.06) ; \draw [shift={(171.6,87.98)}, rotate = 253.67] [color={rgb, 255:red, 0; green, 0; blue, 0 }  ][line width=0.75]    (10.93,-3.29) .. controls (6.95,-1.4) and (3.31,-0.3) .. (0,0) .. controls (3.31,0.3) and (6.95,1.4) .. (10.93,3.29)   ; 
        
        \draw (153.15,20.19) node   [align=left] {$\displaystyle \tau _{2}$};
        \draw (161.65,179.19) node   [align=left] {$\displaystyle \tau _{1}$};
        \draw (426.39,21.19) node   [align=left] {$\displaystyle s_{2}$};
        \draw (427.39,190.19) node   [align=left] {$\displaystyle s_{1}$};
        \draw (360.04,63.71) node   [align=left] {$\displaystyle \beta_{\text{fake}} = \beta/v$};
        \draw (141.2,63.71) node   [align=left] {$\displaystyle \beta $};
    \end{tikzpicture}
    \caption{The fake circle.}
    \label{fig:Fake_circle}
\end{figure}

As observed in \cite{Streicher:2019wek,Lin:2023trc}, this looks like the conformal two point function, $\sin^{-2\Delta}(2\pi s_{12}/\beta_{\text{fake}})$ when written not in terms of the coordinates on the thermal circle, but rather on a ``fake'' circle of circumference $\beta_{\text{fake}} = \beta / v$, \autoref{fig:Fake_circle}, via the mapping 
\begin{equation}
    s_{1}=\tau_{1}+\frac{\beta-\beta_{\text{fake}}}{4}\,,\qquad s_{2}=\tau_{2}-\frac{\beta-\beta_{\text{fake}}}{4} \,, \qquad s_{12} = s_1 - s_2\,.
\end{equation}
where $\tau_1 = \frac{\beta - \tau}{2}$, $\tau_2 = \frac{\beta + \tau}{2}$ are the Euclidean times in which the operators are inserted. The two point function \eqref{eq:finite temperature 2 pt} does not diverge when $\tau_1 = \tau_2$, as the probes annihilate. In the fake disk, the divergence is evaded as the operators are mapped to different points.

Another manifestation of the fake temperature appears in a Lorenzian two-sided correlator, where $\tau_1 = it$, $\tau_2 = \beta/2 - it$. Usually in holographic systems its decay rate is proportional to the temperature, dictated by the quasinormal modes of the dual black hole. However here $G_\Delta \sim \cosh^{-2\Delta}(2\pi t/\beta_{\text{fake}})$, and the rate is proportional to the fake temperature \cite{Lin:2023trc,Blommaert:2024ydx,Rahman:2024vyg}. At high temperatures the decay rate approaches a constant, $\beta_{\text{fake}} \to 1/\bJ$ (see \cite{Lin:2022nss}).

This form of the two point function is protected by a symmetry that originates at finite $\lambda$ \cite{Lin:2023trc}, and in the limit $\lambda \to 0$ becomes an $\fsl_2$ symmetry that acts naturally on the fake circle. We will come back to this in Section~\ref{sec:chord algebra}.

The fake temperature also appears in the expectation value of $\hat n$ in some TFD state\footnote{See \cite{Okuyama:2022szh} for the finite $\lambda$ distribution.}, i.e. in its thermal expectation value. Since the gravitational interpretation of the state is a two-sided geometry, this measures the length $\ell = \lambda \hat n$ of the wormhole connecting the two sides. It is computed by taking derivatives of the two point function \eqref{eq:2 pt function transition amp}
\begin{equation}
    \label{eq:average num of chords}
    \left\langle \text{TFD}| \hat n | \text{TFD} \right\rangle = -\frac{1}{Z}\frac{1}{\lambda}\frac{\partial}{\partial\Delta} G_\Delta\left(\tfrac{\beta}{2},\tfrac{\beta}{2}\right) \Big|_{\Delta = 0} = -\frac{2}{\lambda} \log\left(\beta_{\text{fake}} \cdot \bJ / \pi \right) \,.
\end{equation}
In the triple scaling limit, when real and fake temperatures are the same and $\beta\bJ \propto 1/\lambda$, the renormalized length $\tilde \ell = \ell + 2\log \lambda$ is finite, as we previously claimed. On the other hand, at high temperatures the typical number of chords shrinks to zero \cite{Susskind:2023hnj}.

\paragraph{The chaos exponent}
The four point function, reviewed in Section~\ref{sec:four_point_func}, is more difficult to tackle in the semi-classical limit. We are especially interested in the out-of-time-order correlator, whose connected part is
\begin{equation}
    \label{eq:submaximal_chaos_exp}
    \big\langle \Tr \big[ e^{-\beta H} M_1(0) M_2(it)  M_1(0) M_2(it) \big] 
    \big\rangle_{J,\tilde J_1, \tilde J_2}^{\text{connected}} \propto \lambda \cdot e^{\lambda_L t} \,, \quad \lambda_L = \frac{2\pi v}{\beta} = \frac{2\pi}{\beta_{\text{fake}}}\,,
\end{equation}
The Lyapunov exponent, $\lambda_L$, was computed in the large $p$ limit of SYK \cite{Maldacena:2016hyu} to be smaller\footnote{Actually, with respect to the bound derived in \cite{Parker:2018yvk}, the Lyapunov exponent here is still maximal.} than the maximal value \cite{Maldacena:2015waa} $\lambda_{max} = \frac{2\pi}{\beta}$. The leading deviation from $\lambda_{max}$ at low temperatures was computed directly from the exact double scaled four point function \eqref{eq:OTO_in_terms_of_R} \cite{Berkooz:2018jqr}. The expression for generic $v$ was obtained by \cite{Lin:2023trc} based on the $\fsl_2$ algebra reviewed in Section~\ref{sec:chord algebra}. The algebra also implies the four point function has a simple closed form for any $v$, reproducing the large $p$ result of \cite{Streicher:2019wek, Choi:2019bmd}.


\section{The four point function and  $SU_{\hat{q}}(1,1)$} \label{sec:four_point_func}

Earlier we obtained JT gravity in the $\lambda\rightarrow 0$ and low energy limit. However, the main test is to compute the 4-pt function and chaos exponent. Then we can ask what that means for gravity at finite $\lambda$, which we suggested to be the quantum gravity regime.

\begin{figure}[t]
\centering
    \begin{subfigure}[b]{0.44\textwidth}
        \centering
        \includegraphics[width=\textwidth]{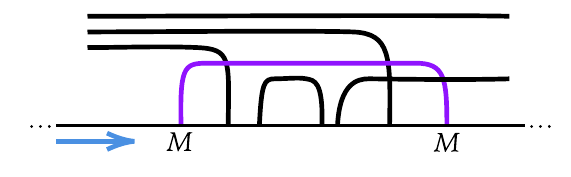}
        \subcaption{A region between contracted $M$ nodes.}
        \label{fig:propagation_M_region}    
    \end{subfigure}
    \hfill
    \begin{subfigure}[b]{0.50\textwidth}
    \centering
        \includegraphics[width=\textwidth]{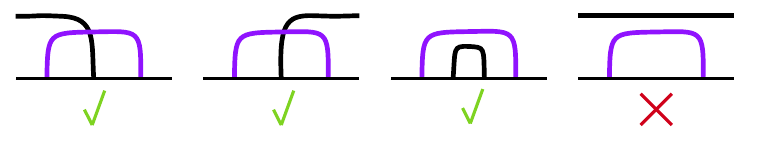}
        \subcaption{
        The last case is incorrectly assigned a value of $\tilde q^2$ when naively inserting $\tilde q^{\,\hat n}$ at $M$ nodes.}
        \label{fig:counting_correlators}    
    \end{subfigure}
    \caption{Computation of the weight of bi-local operators.}
\end{figure}

Our goal is to compute the four point function \cite{Berkooz:2018jqr}. We concentrate on the crossed ordering 
\begin{equation}\label{n-pt sec 3}
G(\tau_1,\tau_3;\tau_2,\tau_4) = \big\langle \Tr  \big(e^{-\beta H} \wick{\c3 M_1(\tau_1) \c2 M_2(\tau_2) \c3 M_1(\tau_3) \c2 M_2(t_4)} \big) \big\rangle_{J,\tilde J_1, \tilde J_2} \,.
\end{equation}
The technical novelty is that even if we ``cut open'' the circular diagram at an insertion point of one of the operators, say $M_1$, as we did for the computation of the two point function, we still need to deal with the double insertion of $M_2$ at two arbitrary point as in figure \autoref{fig:propagation_M_region}. This is the analogue of the bi-local operators in the Schwarzian theory \cite{Maldacena:2016upp}.

For starters, we expect that at each of the insertions of $M_i$ we will need to count the number of chords, so we expect that at each $M_i$ insertion we will need to insert $\tilde q_i^{\,\hat n}$, where $\hat n$ counts the number of open Hamiltonian chords as before (and ${\tilde q}_i$ are the weights associated with $M_i$). 
However, if we just insert this matrix at the two point we will be giving a wrong factor ${\tilde q_i}^{\,2}$ to chords that enter and leave the region without intersecting the $M_i$-chord, see \autoref{fig:counting_correlators}.
We can correct the error by adding $(1 - \tilde q_i^2)$ multiplied by the weight of $n-1$ chords entering the region. But we still pick up an error when 2 chords remain above the $M_i$-chord, which we will need to correct again by the weight of $n - 2$ chords entering the region, and so on. Since $n$ is finite this is a finite set of corrections.

Going from $l$ chords to $l-1$ chords and vice versa, and adding factors of $\tilde q_i$, are conveniently implemented using the matrices
\begin{equation}
U = \begin{pmatrix}
0 & 1 & 0 & \cdots \\
0 & 0 & 1 & \cdots \\
0 & 0 & 0 & \ddots \\
\vdots & \vdots & \ddots & \ddots
\end{pmatrix},\quad 
D = \begin{pmatrix}
0 & 0 & 0 & \cdots \\
1 & 0 & 0 & \cdots \\
0 & 1 & 0 & \ddots \\
\vdots & \vdots & \ddots & \ddots
\end{pmatrix},\quad
S_i = \begin{pmatrix}
1 & 0 & 0 & \cdots \\
0 & \tilde q_i & 0 & \cdots \\
0 & 0 & \tilde q_i^2 & \ddots \\
\vdots & \vdots & \ddots & \ddots
\end{pmatrix} = \tilde q_i^{\,\hat n}
\end{equation}
For $l$ incoming chords into the $M_i$-region consisting of $k$ Hamiltonian nodes, the transition matrix through the entire $M_i$-region can be written as
\begin{equation}\label{eq:propagation_M_region_preliminary}
S_i T^k S_i + P^{(l)}_1 D S_i T^k S_i U +  \cdots + P^{(l)}_l D^l S_i T^k S_i U^l \,, \quad P_{j,i}^{(l)} = \frac{(q;q)_l (\tilde q_i^{\,2};q)_j}{(q;q)_{l-j} (q;q)_j},\quad P_{0,i}^{(l)}=1\,.
\end{equation}
where $P^{(l)}_j$ guarantee there we are not overcounting or undercounting of the diagrams \cite{Berkooz:2018jqr}.
This bi-local operator can be inserted in any correlation function, but specifically for the crossed 4-pt function it gives
\begin{multline} \label{eq:crossed_4_pt_initial}
\langle \Tr \big( H^{k_4}
\contraction[2ex]{}{M_2}{H^{k_3} M_1 H^{k_2} }{M_2}
\contraction{M_2 H^{k_3}}{M_1}{H^{k_2} M_2 H^{k_1}}{M_1}
M_2 H^{k_3} M_1 H^{k_2} M_2 H^{k_1} M_1 \big) \big\rangle_{J,\tilde J_1, \tilde J_2} = \\
= \sum _{l,m,n=0} ^{\infty } \sum _{j=0} ^l P^{(l)} _{j,2} \tilde q_1^{\,j} \langle 0 | T^{k_4} | m\rangle \langle m| D^j S_2 T^{k_3} S_1 |n \rangle \langle n | T^{k_2} S_2 U^j | l \rangle \langle l | T^{k_1}  | 0 \rangle \,.
\end{multline}
After some algebraic manipulation, the crossed four point function becomes \cite{Berkooz:2018jqr}
\begin{multline}
\label{eq:OTO_in_terms_of_R}
G(\tau_1,\tau_3;\tau_2,\tau_4) = \int _0^{\pi } \prod _{j=1} ^4 \big[ d\theta_j \rho(\theta_j) \big] e^{-\beta _1 E(\theta _1) - \beta _2E(\theta _2) -\beta _3 E(\theta _3) -\beta _4 E(\theta _4)} \\
\times \langle \theta_1| M_1 | \theta_4\rangle \langle \theta_3| M_2 | \theta_2\rangle \langle \theta_2| M_1 | \theta_1\rangle \langle \theta_4| M_2 | \theta_3\rangle \times R^{(q)}_{\theta _4\theta _2} \begin{bmatrix}\theta _3 & l_2 \\ \theta _1 & l_1 \end{bmatrix} \, ,
\end{multline}
where we denoted $\beta_1 = \beta + \tau_{41}$, $\beta_2 = \tau_{12}$, $\beta_3 = \tau_{23}$, $\beta_4 = \tau_{34}$, $\tau_{ij} = \tau_i - \tau_j$, and $\tilde q_i = q^{\,l_i}$. The matrix elements already appeared implicitly in the computation of the 2-pt function in Section~\ref{sec:HamChords} and are given explicitly in equation \eqref{eq:diagrammatic_rules_vertex}. The $R$-matrix is defined in \eqref{eq:R_matrix_full_expr}, and is the $6j$-symbol of $SU_q(1,1)$, as we will discuss more below.

\subsection{The diagrammatic rules of the full model} \label{sec:diag_rules}

\begin{figure}[t!]
    \centering
    \begin{subfigure}[t]{0.45\textwidth}
          \centering
          \includegraphics[height=0.4\textwidth]{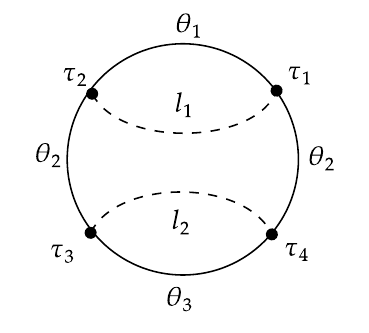}
          \caption{Uncrossed.}
          \label{fig:diagrams_4_point_function_ordered}
    \end{subfigure}
    \quad
    \begin{subfigure}[t]{0.45\textwidth}
          \centering
          \includegraphics[height=0.4\textwidth]{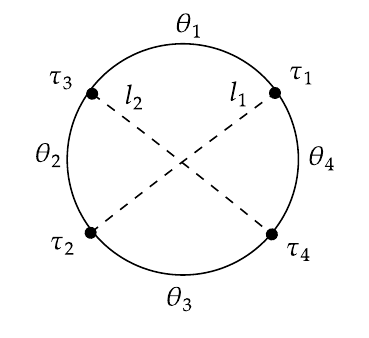}
          \caption{Crossed.}
          \label{fig:diagrams_4_point_function_OTO}
    \end{subfigure}
    \caption{Skeleton diagrams of the 4-point functions $\langle M_1M_1M_2M_2\rangle$ and $\langle M_1 M_2 M_1 M_2\rangle$.}
\end{figure}

While the expressions above look complicated, they are naturally organized in terms of ``skeleton digrams" that keep only the information about the probe operators, and encode the sum over the insertions of the Hamiltonians \cite{Berkooz:2018jqr}. Their diagrammatic rules allow for the calculation of general correlation functions, analogously to the Schwarzian theory  \cite{Mertens:2017mtv, Mertens:2018fds, Lam:2018pvp, Blommaert:2018oro}:
\begin{itemize}
\item Each segment along the circle represents an evolution using $T$ during a Euclidean time $\tau$, and we therefore associate with it a factor of $e^{-\tau  E(\theta)}$. Summation over energy eigenstates is done with the measure $d\theta\rho(\theta)$.
\begin{equation} \label{eq:diagrammatic_rules_propagator}
\eqgraph{6.5ex}{0ex}{\begin{fmffile}{prop}
  \begin{fmfgraph*}(50,0)
    \fmfleft{i1}
    \fmfright{o1}
    \fmfdot{i1,o1}
    \fmflabel{$\tau_1$}{i1}
    \fmflabel{$\tau_2$}{o1}
    \fmf{plain,left,label=$\theta$}{i1,o1}
  \end{fmfgraph*}\end{fmffile}} \qquad = e^{-(\tau_2-\tau_1)  \cdot E(\theta )} .
\end{equation}
\item Each operator insertion between two segments $\theta _1$, $\theta_2$ is naturally thought of as a matrix element $\langle \theta_1 | M_A | \theta_2 \rangle$. In order to reproduce the two point function \eqref{2PTTheta}, the corresponding vertex is
\begin{equation} \label{eq:diagrammatic_rules_vertex}
\begin{gathered}
\begin{fmffile}{vertex}
  \begin{fmfgraph*}(50,50)
    \fmfleft{i1}
    \fmfright{o1,o2,o3}
    \fmfdot{o2}
    \fmf{dashes,label=$l_A$}{o2,i1}
    \fmf{plain,right=0.1,label=$\theta _2$}{o1,o2}
    \fmf{plain,right=0.1,label=$\theta _1$}{o2,o3}
  \end{fmfgraph*}\end{fmffile}
\end{gathered}  \qquad = \langle \theta_1 | M_A | \theta_2 \rangle = \sqrt{\frac{(\tilde q_A^2;q)_{\infty } }{\left( \tilde q_A e^{i(\pm \theta _1 \pm \theta _2)} ;q\right)_{\infty } } } \,,
\end{equation}
\item A contracted pair of operators conserves the energy before and after the insertion \cite{Berkooz:2018jqr}. Thus, the same $\theta $ variable is used before and after such a contracted pair if there is no additional operator insertions in between, as in \autoref{fig:diagrams_4_point_function_ordered}, 
\begin{equation} \label{eq:diagrammatic_rules_en_cons}
\begin{gathered}
\begin{fmffile}{en_cons}
  \begin{fmfgraph*}(50,50)
    \fmfleft{i1,i2,i3}
    \fmfright{o1,o2,o3}
    \fmfdot{i2,o2}
    \fmf{dashes,label=$l_A$,label.side=left}{o2,i2}
    \fmf{plain,right=0.2,label=$\theta_2$}{o1,o2}
    \fmf{plain,right=0.7,label=$\theta_1$}{o2,i2}
    \fmf{plain,right=0.2,label=$\theta_2$}{i2,i1}
  \end{fmfgraph*}\end{fmffile}
\end{gathered}  \qquad .
\end{equation}
\item Every crossing of two internal lines comes with a factor of the $R$-matrix, \eqref{eq:R_matrix_full_expr},
\begin{equation} \label{eq:diagrammatic_rules_crossing}
    \vcenter{\hbox{\begin{tikzpicture}[x=0.75pt,y=0.75pt,yscale=-0.7,xscale=1]
        \draw  [dash pattern={on 4.5pt off 4.5pt}]  (45,141) .. controls (90,105.5) and (125,79.5) .. (163,51.5) ;
        \draw  [dash pattern={on 4.5pt off 4.5pt}]  (46,52) .. controls (78,75.5) and (130,116.5) .. (161,142) ;
        
        \draw (133,42) node [anchor=north west][inner sep=0.75pt]   [align=left] {$\displaystyle l_{1}$};
        \draw (66,44) node [anchor=north west][inner sep=0.75pt]   [align=left] {$\displaystyle l_{2}$};
        \draw (95,16) node [anchor=north west][inner sep=0.75pt]   [align=left] {$\displaystyle \theta _{1}$};
        \draw (167,84) node [anchor=north west][inner sep=0.75pt]   [align=left] {$\displaystyle \theta _{4}$};
        \draw (94,146) node [anchor=north west][inner sep=0.75pt]   [align=left] {$\displaystyle \theta _{3}$};
        \draw (26,86) node [anchor=north west][inner sep=0.75pt]   [align=left] {$\displaystyle \theta _{2}$};   
    \end{tikzpicture}}} 
    \quad = \ 
    R^{(q)}_{\theta _4\theta _2} \begin{bmatrix}\theta _3 & l_2 \\ \theta _1 & l_1 \end{bmatrix} \times 
    \vcenter{\hbox{\begin{tikzpicture}[x=0.75pt,y=0.75pt,yscale=-0.7,xscale=1]
        \draw  [dash pattern={on 4.5pt off 4.5pt}]  (41,149) .. controls (58,114) and (140,114) .. (157,150) ;
        \draw  [dash pattern={on 4.5pt off 4.5pt}]  (42,60) .. controls (60,95) and (147,93) .. (160,58) ;
        \draw (94,57) node [anchor=north west][inner sep=0.75pt]   [align=left] {$\displaystyle l_{1}$};
        \draw (94,132) node [anchor=north west][inner sep=0.75pt]   [align=left] {$\displaystyle l_{2}$};
        \draw (94,23) node [anchor=north west][inner sep=0.75pt]   [align=left] {$\displaystyle \theta _{1}$};
        \draw (155,91) node [anchor=north west][inner sep=0.75pt]   [align=left] {$\displaystyle \theta _{2}$};
        \draw (94,164) node [anchor=north west][inner sep=0.75pt]   [align=left] {$\displaystyle \theta _{3}$};
        \draw (30,91) node [anchor=north west][inner sep=0.75pt]   [align=left] {$\displaystyle \theta _{2}$};
    \end{tikzpicture}}}
    \,,
\end{equation}
\end{itemize}
Applying these rules to the crossed diagram Figure~\ref{fig:diagrams_4_point_function_OTO}, reproduces the explicit result \eqref{eq:OTO_in_terms_of_R}.

In the triple scaling limit of Section~\ref{sec:triple scaling}, these diagrammatic rules become those of the Schwarzian theory, described in \cite{Mertens:2017mtv,Mertens:2018fds}. Yet, the rules described here are exact at all energy scales and for any $q = e^{-\lambda}$, not just semiclassically.

In the finite $p$ SYK the connected 4-pt function from which the chaos exponent is extracted is suppressed by $1/N$, reflecting the fact that the interaction is gravitational. Here the expression is finite, and goes to zero when $\lambda\rightarrow 0$ supporting the idea that the theory has a finite Planck scale, controlled by $\lambda$.

Finally, the core of the 4-pt function is the $R$-matrix. In the Schwarzian case, it is the $6j$ symbol of $SU(1,1)$ \cite{Mertens:2017mtv,Mertens:2018fds,Blommaert:2018oro,Iliesiu:2019xuh,Suh:2020lco}. In our case, the $R$-matrix is the $6j$ symbol of the quantum group $SU_{\sqrt{q}}(1,1)$ \cite{Berkooz:2018jqr,groenevelt2006wilson}. In addition, the chord wavefunctions \eqref{eq:energy eigenstates} are representation matrix elements\footnote{A different $q$ deformation of $\fsl_2$, $SL_q(2,\bR)$, is exhibited by Liouville gravity \cite{Blommaert:2023wad,Mertens:2022aou,Fan:2021bwt}.} of $SU_{\sqrt{q}}(1,1)$, and the vertex \eqref{eq:diagrammatic_rules_vertex} originates from their Clebsch-Gordan coefficients \cite{Blommaert:2023opb}. This is the key to the non-commutative geometry structure of the model, which we turn to now.


\section{Non-commutative AdS (bulk 2)}\label{sec:non-commutative}

Just like groups describe the symmetries of usual geometric spaces, quantum groups capture the symmetries of non-commutative geometries. 
So far, we have seen two appearances of objects from the realm of non-commutative geometry. First, the $q$-oscillator algebra \eqref{eq:ArikCoon}. Second, the four point function is related to the $6j$ symbol of the quantum group $SU_{\hq}(1,1)$, which parallels a similar role as the Lie group $SU(1,1)$ in the Schwarzian theory. But the latter also controls the full geometry, so one then wonders whether $SU_{\hq}(1,1)$ can be related to the full geometry also for DS-SYK, and in particular whether non-commutative geometry replaces ordinary geometry in the bulk. The answer seems to be yes, with some guesswork. 

The main points of Section~\ref{sec:non-comm_ads2} is that one can construct a set of non-commutative $AdS_2$ spaces as homogeneous spaces under the action of $SU_{\hat q}(1,1)$. The transfer matrix is then a reduction of a Casimir for this action, and appears as the Hamiltonian of a particle on this non-commutative space \cite{Berkooz:2022mfk,Blommaert:2023opb}. This generalizes the construction of Schwarzian as the action of the boundary particle moving on $AdS_2$. 

For proper disclosure -- the picture is not complete. One immediate problem is that there is more than one $q$-deformed $AdS_2$, so in any approach one needs to add some more assumptions in order the get ``the correct'' one. However, the main problem in our opinion is that it is not clear what is the microscopic origin of any of the non-commutative geometry objects that we will discuss\footnote {of course still after ensemble average and not realization-by-realization.}.

 In Section~\ref{sec:chord algebra} we review the  generalization of the $q$-oscillator algebra in the presence of matter chords \cite{Lin:2022nss}, and highlight a specific subalgebra that generalizes the global $\fsu(1,1)$ symmetry of the Schwarzian theory when coupled to matter. The semi-classical limit of this algebra is the origin of the fake temperature of Section~\ref{sec:semi-classics} \cite{Lin:2023trc}.

But first, we would like to provide a (brief and non-systematic) definition of the quantum algebra $\cU_{\hat q}(\mathfrak{s}\mathfrak{u}(1,1))$ and its associated quantum group $SU_{\hq}(1,1)$. In order to conform with the standard notation on quantum groups, we define
\begin{equation}
    {\hat q} = q^{1\over 2} = e^{-\frac{1}{2}\lambda} \,.
\end{equation}
The algebra\footnote{Which can also be thought of as a deformation of the universal enveloping algebra $\cU(\fsu(1,1))$.} $\mathcal{U}_{\hat q}(\mathfrak{s}\mathfrak{u}(1,1))$ is generated by elements $\cA$, $\cB$, $\cC$, $\cD$, modulo the following relations:
\begin{equation}\label{SLq2}
    \cA \cD = \cD \cA = 1, \qquad \cA \cB = {\hat q} \cB \cA, \qquad \cA \cC  = {\hat q}^{-1} \cC \cA,\qquad [\cB,\cC] = {\cA^2-\cD^2\over {\hat q} -{\hat q}^{-1}}\,.
\end{equation}
We could have not introduced $\mathcal{D}$ and just used ${\mathcal A}^{-1}$, however, written as an algebra of symbols modulo relation this way is a little more precise. In addition, not every symbol in the algebra has an inverse.

The Casimir element, generating the center of this algebra, is given by 
\begin{equation}
        \Omega  = \left(  { {\hat q}^{-1}\cA^2 + {\hat q} \cD^2-2 \over ({\hat q}^{-1} - {\hat q})^2} + \cB \cC \right)\,.
\end{equation}
In the limit $\hq \to 1$, formally defining $\cA = \hq^{\cE}$, the algebra reduces to $\fsu(1,1)$ as $[\cE,\cB] = \cB$, $[\cE,\cC] = -\cC$, $[\cB,\cC] = -2\cE$.

One way \cite{Blommaert:2023opb}, which will be used below, to write elements of the quantum group $SU_{\hq}(1,1)$ in terms of these generators is by a Gauss decomposition using non-commutative coordinates,
\begin{equation}
\label{eq:non-commutative coords}
    e_{{\hq}^{-2}}^{\gamma \cC} \cA^{2\phi} e_{{\hq}^2}^{\beta \cB} \in SU_{\hq}(1,1) \,, \qquad \left[\gamma,\phi\right] = \frac{\lambda}{2}\gamma\,, \qquad\left[\beta,\phi\right] = \frac{\lambda}{2}\beta\,, \qquad \left[\beta,\gamma\right] = 0 \,.
\end{equation}
where $e_q^x$ is the $q$-exponential \eqref{eq:q-exponential def} -- a $q$-deformation of the exponential function, motivated by the action of the $q$-derivative $D_q e_q^z\equiv {e_q(qz)-e_q(z)\over qz-z}=e_q^z$ and by $e_q(z/(1-q))\rightarrow e^z$.

\subsection{Non-commutative $AdS_2$}
\label{sec:non-comm_ads2}

In the triple scaling limit the transfer matrix reduces to the Liouville form of the Schwarzian theory \cite{Bagrets:2017pwq}. There are many different formulations of the latter -- from JT gravity on the disk topology, to a BF theory with an $\fsl_2(\bR)$ gauge algebra on that space\footnote{Even though this formulation is not precisely complete because of issues with the correct boundary conditions and the global form of the gauge group.}. All formulations reduce to a boundary action, which is that of a particle moving on $AdS_2$ (with an effective large imaginary magnetic field). 

Since we lack a rigorous analogue for the gravitational or BF path integral for DS-SYK (see however recent advances in Section~\ref{sec:Gravitational path integral}), we  aim for a generalization of the boundary particle action. The key point is that $AdS_2$ is an homogeneous space of $SL(2,\bR)$, and the Hamiltonian which controls the motion on it is the Casimir of the group on an appropriate set of representations. In the context of DS-SYK, the transfer matrix that we constructed above will turn out to be a Casimir of $SU_{\hat q}(1,1)$ on such a set.

We will briefly go through the construction for the case of ordinary $AdS_2$ mainly because we will see the same components in the non-commutative version, in an almost one to one fashion. The reduction from the motion of the boundary particles on $AdS_2$ to the Schwarzian action is encoded in the formula  $Z(\beta)=\langle X| e^{-\beta H_{AdS_2}} | X\rangle$
where $Z(\beta)$ is the partition function of the SYK model at low energies, or equivalently the Schwarzian action  \cite{Kitaev:2018wpr} (see also \cite{Mertens:2017mtv}). On the RHS we have the propagator of the boundary particle from point X back to X, and $H_{AdS_2}$ is its Hamiltonian (which is a particle moving on $AdS_2$ with a large imaginary magnetic field). $X$ is any point on $AdS_2$, and all choices of $X$'s are gauge equivalent. 

In Euclidean Poincare coordinates a particle moving on $AdS_2$ with coordinates $ds^2 = d\phi^2 + e^{2\phi} df^2$ in the presence of a large magentic field is given by \cite{Mertens:2017mtv}
\begin{equation}\label{eq:AdS2LrgMag}
    S=\int dt \left( {1\over 4} {\dot\phi}^2 +{1\over 4}e^{2\phi}{\dot f}^2 + B {\dot f}e^{\phi}\right) \,.
\end{equation}
Introducing the canonical momenta $\pi_f, \pi_\phi$, the Hamiltonian for the system is 
\begin{equation}
\label{eq:Liouville Hamiltonian mag field}
H_B = \pi_\phi^2 + (\pi_f e^{-\phi} - B)^2 \longrightarrow \pi_\phi^2 + e^{-\phi} \,,
\end{equation}
where on the right side we fixed $\pi_f$ (which is an invariant of the motion), shifted $\phi$ by $ -log(-2B\pi_f)$ and took the $B \to \infty$ limit, resulting in the Liouville form of the Schwarzian. The $SL(2,\bR)$ action on this $AdS_2$ is a composition of spatial translation and a gauge symmetry
\begin{equation}\label{eq:sl2Gen}
l_{-1} = \pi_f\,, \quad l_0 = f \pi_f - \pi_\phi\, \quad l_1 = f^2 \pi_f - 2f \pi_\phi - \pi_f e^{-2\phi} + 2Be^{-\phi} \,,
\end{equation}
and $H$ is just the Casimir of this geometric $SL(2)$. We will next see the non-commutative versions of these expressions.

\paragraph{Some ad-hoc realizations}

There could be many different\footnote{
It is worth mentioning a different non-commutative space which is the analogue of the Euclidean disk with rotationally invariant coordinates, $\frac{dz\otimes dz^*}{(1-zz^*)^2}$, which is natural in holography. The non-commutative analog was analyzed in \cite{Almheiri:2024ayc}, focusing on wave functions of ordinary particles on this space, but not on the boundary particle. This would be necessary next step to connect with the transfer matrix.} realizations of non-commutative $AdS_2$. 
We are going to begin with some heuristics which explains the basic structure and then switch to the more canonical non-commutative geometry point of view. 

Non-commutative spaces are usually defined by deforming the algebra of functions on the space. So to define a non-commutative $AdS_2$ as a homogeneous space of \eqref{SLq2}, we are interested in deforming functions of two variables on which 1) we realize this algebra and 2) have an $AdS_2$ limit when $\hq\rightarrow 1$. Sometimes, however, one can get away with simpler constructions, such as functions on a 2D lattice. 
The lattice appears here naturally because ${\cal A}$ is a group like element and not an infinitesimal generator (it's roughly $\hq^{\,l_0}$), and we expect that ${\cal A}$ will act discretely on a lattice of points. This actually turns all the generators to ones that act on a lattice of points. If we define the discrete translation operators $\cT$, $\cR$,
\begin{equation}\label{eq:def T and R}
    \cT \ F(\tilde H,\tilde z) =  F(\hq^{1 \over 2} \tilde H,\tilde z) \,, \qquad \cR \ F(\tilde H,\tilde z) =  F(\tilde H,\hq \tilde z)\,,
\end{equation}
where $F$ is a function on lattice of points $
({\tilde H},{\tilde z})\in \left\{ (\hq^{m_1/2},\pm\hq^{m_2})\, | \, m_1,m_2\in \mathbb{Z} \right\} \equiv \mathbb{R}^2_{\hq} \subset \mathbb{R}^2 \,,$ then the following generators satisfy \eqref{SLq2} for any integers $a$, ${\tilde a}$
\begin{equation}\label{eq:HandOn}
    \begin{aligned}
        \mathcal{A} & =  \cT \cR^{-1} \,,
        \quad \mathcal{B} = \frac{1}{\tilde z} {\cR^{a+2} - \cR^a \over \hq^{a+2} - \hq^a} \,, 
        \quad \mathcal{C} = {\tilde z \over \hq^{-1} - \hq} \left(  \cT^{-2} \cR^{-a} -   \cT^2 \cR^{-a-2}  \right) + i \mu \tilde H^{-2} \cT^{\tilde a}\,, \\
        \Omega    & \equiv \left(  { \hq^{-1}\mathcal{A}^2 + \hq \mathcal{D}^2-2 \over (\hq^{-1} - \hq)^2} + \mathcal{B}  \mathcal{C} \right) 
        =  {\hq^{-1} \cT^{-2}  + \hq \cT^2-2 \over (\hq^{-1}-\hq)^2  } + {i \mu  \tilde H^{-2} }\cT^{\tilde a} \mathcal{B} \,,
    \end{aligned}
\end{equation}
where we have also written the Casimir $\Omega$. In the limit $\hq \to 1$, the $AdS_2$ coordinates are identified as $\tilde H = e^{\phi}$ and $\tilde z = f$, and the translation operators formally become $\cT = e^{\lambda \partial_\phi}$ and $\cR = e^{-\lambda f \partial_f}$, such that $\cA$, $\cB$, and $\cC$ become\footnote{In the limit where $B = i\mu$ is large. The canonical momenta $\pi_{f}$, $\pi_\phi$ act as derivatives $\partial_{f}$, $\partial_\phi$.} $\hq^{l_0}$, $l_{-1}$, $l_1$ of \eqref{eq:Liouville Hamiltonian mag field}.

To go from $\Omega$ in the \eqref{eq:HandOn} to the transfer matrix we mimic the steps around \eqref{eq:Liouville Hamiltonian mag field}. There we fixed $\pi_f$, the eigenvalue of $l_{-1}$, which in \eqref{eq:HandOn} corresponds to the operator $\cB$. Its eigenvectors are some $q$-deformations of the exponential function\footnote{However, orthogonality and completeness relations are not known for most values of $a$.} that depend on $a$. For example for $a=0$ these are $e_{{\hat q}^2}^{c \tilde z}$. Once we digonalize $\cB$ we are left with a Casimir which only has $\cT$ and $\tilde H$ -- i.e., we reduced $\Omega$ to a finite difference operator on the discrete lattice of ${\tilde H}$. By properly choosing the eigenvalue of $\cB$, as well as $\mu$ (which plays the role of the magnetic field), and also setting ${\tilde a}=2$, it is easy to see that we obtain the DS-SYK transfer matrix \eqref{eq:DefChordCreation} (after a simple conjugation) acting on the discrete lattice of ${\tilde H}$.

So we can construct realizations of the quantum group which reproduce the correct transfer matrix and go over to $AdS_2$ when $q\rightarrow 1$. In this approach, however, we had to do some tinkering and guesswork\footnote{Although it could be that different realizations are actually the same under complicated conjugation.}. Below we will see other types of constructions but they will also, by and large, suffer from similar (and partially overlapping) arbitrariness.

\paragraph{Non-commuting coordinates}

One way of obtaining \eqref{eq:AdS2LrgMag} is to formally start with a Minkowskian $AdS_3$ and take $B$ to be the momenta along one of the null coordinates, leaving us with the other null coordinate, which is $f$, and with the radial coordinate $\phi$ \cite{Mertens:2017mtv}. So a possible starting point for DS-SYK is to $q$-deform this $AdS_3$, then reduce to a 2D space, and then to a transfer matrix acting on a 1D lattice. We will denote the commuting $AdS_3$ by
\begin{align}
ds^2_{\mathbb{H}^3} = {dH^2 \over H^2} + H^2 dz dz^* \,.
\end{align}
After the reduction we identify $z\rightarrow f$ and $H\rightarrow e^{\phi}$ to match with \eqref{eq:AdS2LrgMag}.

$H$, $z$, $z^*$ are now promoted to non-commuting variables, and some known non-commuting spaces that generalize $AdS_3$ include \cite{Olshanetsky:1993sw, Olshanetsky:2001, Berkooz:2022mfk} 
\begin{equation}
\begin{aligned}
    {\mathbb{H}}_{\hq}^3:\quad & H=H^*,\quad Hz = \hq^2 z H, \quad z^* H = \hq^2 H z^*, 
    \quad z z^* = \hq^2 z^* z + (\hq^2-1) (1-H^{-2})\,,\\
    {\mathbb{H}}_{\hq, \kappa}^3: \quad &zH=\kappa^{-1}Hz,\quad zz^*=az^*z-b H^{-2},
  \quad a={\kappa^2\over \hq^2}\,,\quad b= {\kappa\over \hq^2} (1-{\hq^2})\,.
\end{aligned}    \label{eq:OR1 Lobachevsky commutations}
\end{equation}
The next step is to define functions on these spaces. The construction starts with polynomials of the non-commuting variables, and then extend it to a richer family of functions. After the deformation the main subtlety is to deal with the ordering of the non-commuting coordinates, but there will be a natural action of $SU_{\hq}(1,1)$ on these functions. We then carry out the reduction to an $AdS_2$ by diagonalizing the appropriate generator giving us a non-commutative $AdS_{2,\hq}$. So far all the non-commutative objects are now in correspondence to those of ordinary $AdS_2$. We then further reduce to a 1D system by diagonlizing the non-commutative analogue of $l_{-1}$ in \eqref{eq:sl2Gen}, as before.

In slightly more detail: a convenient set of normal ordered monomials is
\begin{equation}\label{eq:H-monom-1}
    v^{m,r,n} \equiv   (\hq^{1\over 2}z^*)^{m} H^r(\hq^{1\over 2}z)^{n}\,.
\end{equation}
The reduction to  $AdS_{2,\hq}$, which means for us functions of $H$ and $z$, is done by the ansatz
\begin{equation}
F_{\mu}(z^*,H,z) \colonequals  e_{\hq^2}(i \mu \hq^{1 \over 2 } z^*) \cdot F_\mu(H,z),\quad  F_{\mu}(H,z)= \sum_{r,n}  a_{r+1,n} H^r (\hq^{1 \over 2}z)^n \,, 
\label{eq:Non Commuting AdS3 to AdS2}
\end{equation}
where we kept the ordering of the symbols in $F_\mu$, and the $a_{r+1,n}$ are some constants.

The action of the quantum group on the $q$-deformed $AdS_3$ now descends to an action on $F_\mu$, so these $(H,z)$ have full claim to be considered a $AdS_{2,\hq}$ $q$-deformed homogeneous space. In fact, the generators that one obtains in this approach are very similar to \eqref{eq:HandOn}, except that now the symbols are non-commuting variables and one needs to be careful about their ordering (for complete details see \cite{Berkooz:2022mfk}). The subsequent steps are also very similar. We replace the general $z$ dependence by a $q$-exponential, which is the same as diagonalizing $\cB$, and obtain an action on a discrete lattice of $H$, again with some free parameters that we have to choose correctly. If we now go to the space of eigenvalues of $H$ we get the transfer matrix acting on a discrete lattice.

\paragraph{Particle on a group manifold}

Another way of getting a particle on non-commutative $AdS_3$ is to start with a particle on the quantum group $SU_{\hq}(1,1)$ \cite{Blommaert:2023opb}. This is more compact than the constructions above, and an easier reduction from $AdS_3$ to $AdS_2$, but it only gives one of the options, and still some guesswork is needed. The particle moves in a 6D phase-space 
with coordinates $\gamma$, $\phi$, $\beta$, $\pi_{\gamma}$, $\pi_{\phi}$, $\pi_{\beta}$ and equipped with the symplectic two-form
\begin{equation}
    \omega=d\gamma\wedge d\pi_{\gamma}+d\beta\wedge d\pi_{\beta}+d\left(\phi+i\lambda\left(\gamma\pi_{\gamma}+\beta\pi_{\beta}\right)\right)\wedge d\pi_{\phi}\,,
\end{equation}
i.e. the Darboux coordinates are   $\varphi=\phi+i\lambda\left(\gamma\pi_{\gamma}+\beta\pi_{\beta}\right)$, $\pi_{\varphi}=\pi_{\phi}$. Upon quantization, these satisfy the commutation relations \eqref{eq:non-commutative coords} for the non-commutative coordinates on $SU_{\hat q}\left(1,1\right)$.

In\footnote{Their $q$ is our $\hq$, and therfore we rescaled their $\phi$ by a factor of $2$ and their $\pi_\phi$ by a factor of $1/2$.} \cite{Blommaert:2023opb} the authors constructed left- and right-regular representations of the algebra $\cU_{\hat q}(\fsu(1,1))$ that act on this space, and a Hamiltonian that commutes with them
\begin{equation}
    H = -\frac{1}{\lambda^{2}}\left[e^{-i\lambda\pi_{\phi}}+e^{i\lambda\pi_{\phi}}+e^{-\phi}e^{i\lambda\pi_{\phi}}\frac{\left(e^{-i\lambda\beta\pi_{\beta}}-1\right)\left(e^{-i\lambda\gamma\pi_{\gamma}}-1\right)}{\beta\gamma}\right] \,.
\end{equation}
which is \eqref{eq:HandOn} for the specific values of $a = 0$ and $\tilde a = 2$. In the $\lambda \to 0$ limit we reproduce \eqref{eq:Liouville Hamiltonian mag field} after treating $\pi_\gamma$ as the magnetic field $B \to \infty$, $\pi_\beta$ as $\pi_f$, and shifting $\phi$.

In the quantization of JT gravity a similar picture arises, where additional constraints coming from the boundary conditions in the gravitational theory are imposed. \cite{Blommaert:2023opb} suggested their analog would be
\begin{equation}
    \frac{1}{\gamma}\left(e^{i\lambda\gamma\pi_{\gamma}} -1 \right) e^{\frac{i}{2}\lambda \gamma \pi_\gamma} e^{-\frac{i}{2}\lambda\pi_{\varphi}}=i\,, \quad \frac{1}{\beta}\left(e^{-\frac{i}{2}\lambda\beta\pi_{\beta}}-e^{\frac{i}{2}\lambda\beta\pi_{\beta}}\right)e^{\frac{i}{2}\lambda\pi_{\varphi}} = -i \,.
\end{equation}
After a shift of $\phi$, the Hamiltonian turns into
\begin{equation}
    H=-\frac{1}{\lambda^{2}}\left[\left(1-e^{-\phi}\right)e^{i\lambda\pi_{\phi}}+e^{-i\lambda\pi_{\phi}}\right] \,,
\end{equation}
exactly\footnote{Up to an overall factor, which could be absorbed into the definition of the dimensionful parameter.} as the chord Hamiltonian \eqref{eq:q-Liouville Hamiltonian}. A generalization of the path integral form of the Schwarzian, named the $q$-Schwarzian \cite{Blommaert:2023opb}, can be found as a phase space path integral by similar constraints.

\subsection{The chord algebra}\label{sec:FkBlk}
\label{sec:chord algebra}

Another avatar of the algebraic structure of the theory is the $q$-oscillator algebra of the chords, \eqref{eq:ArikCoon}. This algebraic structure only becomes richer once we consider not only two sided states that contain Hamiltonian chords, as we did in Section~\ref{sec:HamChords}, but also matter chords \cite{Lin:2022rbf, Lin:2023trc}. The advantage of this is that the $q$-deformed SL(2) manifests itself as a global symmetry. In the semi-classical limit, it naturally acts on the ``fake disk'' geometries, and implies a sub-maximal chaos exponent.

\begin{figure}[t]
    \centering
    \begin{subfigure}{0.45\textwidth}
        \centering
        \includegraphics[width=0.5\textwidth]{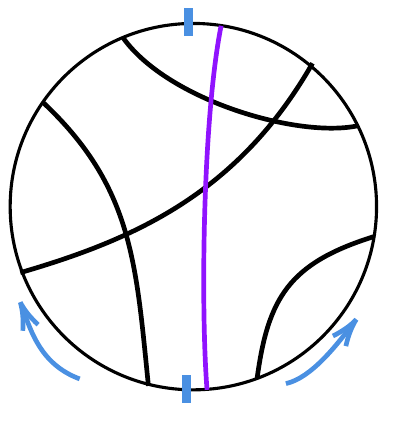}
    \end{subfigure}
    \qquad
    \begin{subfigure}{0.45\textwidth}
        \centering
        \includegraphics[width=0.5\textwidth]{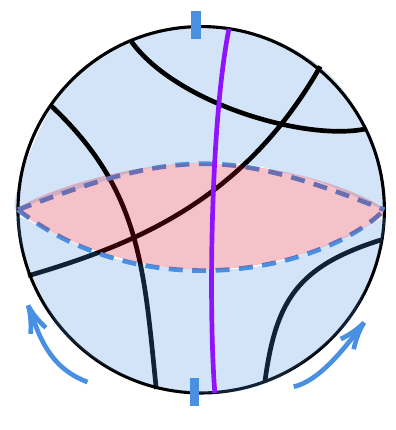}
    \end{subfigure}
    \caption{Cutting the chord diagrams for the 1-particle Transfer matrix.}
    \label{fig:1-particle hilbert space diagams}
\end{figure}

Let's start with the (two-sided) Hilbert space containing a single particle, which corresponds to an operator of dimension $\Delta$, such that an intersection of a Hamiltonian chord with the matter chord results in a factor of $q^\Delta$. We can build all chord diagrams that contribute to the thermal 2-pt function step-by-step, as in the TFD analysis of Section~\ref{sec:HamChords}, by evolving time on the left and on the right. We start by cutting the diagram around the matter chord, \autoref{fig:1-particle hilbert space diagams}. At each step we open (or close) a Hamiltonian chord from one of the boundaries, 
and state of the system is given by the number of chords that are still open to the left ($n$) and to the right ($m$) of the matter chord, $\ket{n, m}$. As before, we choose a convention where chords intersect when they close. We can summarize the evolution by two transfer matrices, describing chords that open and close from the left (right) boundary\footnote{Under a convention where chords intersect when they open, related to ours by conjugation, the 1-particle eigenstates are the bi-variate $q$-Hermite polynomials, \cite{Xu:2024hoc, Okuyama:2024gsn}, up to this conjugation issue.} \cite{Lin:2022rbf},
%
%
%
\begin{align}
    T_L &= a^\dagger_L + a_L \,, & T_R &= a^\dagger_R + a_R \,, \\
    a^\dagger_L \ket{n,m} &= \ket{n+1,m} \,,  &a_L \ket{n,m} &= q^{\Delta + n} [m]_q \ket{n,m-1} + [n]_q \ket{n-1,m} \,, \\
    a^\dagger_R \ket{n,m} &= \ket{n,m+1} \,,   &a_R \ket{n,m} &= q^{\Delta + m} [n]_q \ket{n-1,m} + [m]_q \ket{n,m-1} \,.
\end{align}


As with the TFD state, we can view $e^{\frac{1}{2}(\beta_L T_L + \beta_R T_R)} |0,0\rangle$ as generating all ``bottom halves'' of diagrams with a single matter chord. In order to reproduce the two point function from an inner product of two such states, we will need it to account for the intersections of the chords that leave these two halves, see \autoref{fig:1-particle hilbert space diagams}, 
\begin{equation}
    \braket{n_2,m_2 |n_1,m_1} = \delta_{n_2 + m_2, n_1 + m_1} \sum_{\text{pairings}} q^{\text{no. of $H$-$H$ intersections}} q^{\Delta\times \text{ no. of $H$-$M$ intersections}}\,.
\end{equation}
The inner product\footnote{A closed formula can be found in \cite{Lin:2023trc}.} allows $H$-chords to cross from one side of the matter chord to the other, and the basis $\ket{n,m}$ is not orthogonal. Yet, the creation and annihilation operators $a_{L/R}$ and $a^\dagger_{L/R}$ are Hermitian conjugates, and the inner product is positive definite \cite{Speicher:1993kt,pluma2022dynamical,Berkooz:2020xne}.


States with different $n+m$ are orthogonal, though, and so we define an operator that measures this number, the \emph{size operator}, $\hat n$. Together with the chord creation and annihilation operators it generates the \emph{chord algebra} \cite{Lin:2023trc},
\begin{align}
    \label{eq:chord algebra}
    \hat n \ket{n,m} &= n + \Delta + m \,,  &[\hat n, a_{L,R}] &= -a_{L,R} \,,
    & [\hat n, a^\dagger_{L,R}] &= a^\dagger_{L,R} \,, \\
     [a_L, a_R^\dagger] &= [a_R, a_L^\dagger] = q^{\hat n}  \,,
    &[a_L,a_R] &= [a_L^\dagger, a_R^\dagger] = 0\,,  & [a_L, a_L^\dagger]_q &= [a_R, a_R^\dagger]_q = 1 \,.
\end{align}
which generalizes the $q$-oscillator algebra \eqref{eq:ArikCoon}. In fact, the latter could be thought of as a short representation of the chord algebra in which $a_L = a_R$ and $a_L^\dagger = a_R^\dagger$, and $\Delta = 0$. The only \cite{Lin:2023trc} other highest weight irrep for the algebra is the one with a single matter chord discussed above. There are, however, reducible representations in which there are several matter chords, but only $H$-chords open and close dynamically.

The commutation relations imply $[T_L, T_R] = 0$ and the operatorial equation \cite{Lin:2023trc}
\begin{equation}
    \label{eq:finite q eom}
    \left[T_{L},\left[T_{R},\hat{n}\right]\right] = -2q^{\hat{n}} 
 \quad \implies \quad \partial_{\tau_1}\partial_{\tau_2} \hat n = -2q^{\,\hat n} \,,
\end{equation}
when the evolution along the two boundaries is generated by the respective transfer matrix. This generalizes the evolution using the Liouville Hamiltonian \eqref{eq:triple_scaling_Liouville_Ham} to finite $\lambda$, when $\ell = \lambda \hat n$.

A curious feature is a sub-algebra \cite{Lin:2023trc}, $U(J)$, that commutes with $\hat n$ and is generated by
\begin{equation}
    J_{ij} = a^\dagger_i a_j - [\hat n]_q \,, \qquad i,j\in\{L,R\}\,,
\end{equation}
In Appendix C of \cite{Lin:2023trc} it was shown that it generates a proper subalgebra of $\cU_{\hq}(\mathfrak{sl}_2)$. The $U(J)$ algebra, together with $c = q^{\hat{n}/2}$, can be used to define the generators
\begin{equation}
    B = \frac{1}{2c}\left(J_{LL}-J_{RR}\right) \,, \quad 
    E = -\frac{1}{2c}\left(J_{LL}+J_{RR}\right) \,, \quad   P = \frac{i}{2}\left(J_{LR}-J_{RL}\right) = iq[B,E] \,.
\end{equation}
 When $\lambda \to 0$, \eqref{eq:average num of chords} implies that to leading order $c = \frac{\pi v}{\beta\bJ} = \frac{\pi}{\beta_{\text{fake}}\cdot \bJ}$ is a c-number rather then an operator. The rescaled generators $\mathsf{B} = \frac{B}{\sqrt{1-c^2}}$, $\mathsf{E} = E$, $\mathsf{P} = \frac{P}{\sqrt{1-c^2}}$ then generate an $\fsl_2(\bR)$ algebra. The 1-matter chord representation described above transforms under this algebra, implying its action on two-point functions \cite{Lin:2023trc}. This turns out to be the natural action of $\fsl_2(\bR)$ on a circle with the circumference $\beta_{\text{fake}}$, the fake circle described in Section~\ref{sec:semi-classics}. Moreover, around small perturbations to the thermofield double state $2c\mathsf{B} = T_R - T_L$ as an operatorial equation, implying an exponential growth in Lorentzian time for the operator $\mathsf{P}^- = \mathsf{E} - \mathsf{P}$ (which we will treat as the ``scramblon''),
 \begin{equation}
     e^{i(T_R-T_L)t}\mathsf{P}^-e^{-i(T_R-T_L)t} = e^{2ct\bJ}\mathsf{P}^- \,,
 \end{equation}
and suggesting that the Lypaunov exponent is $\lambda_L = 2c = 2\pi/\beta_{\text{fake}}$ \cite{Lin:2023trc} as in Section~\ref{sec:semi-classics}.


 
\section{Generalizations and variations} \label{sec:Generalizations}

We have seen in Section \ref{sec:HamChords} that the chord diagram technique can be applied to several $p$-local systems in the double scaling limit. Actually, from the discussion there it is clear that it applies to a large class of models. Additional models analyzed so far include the double scaled sparse-SYK \cite{Garcia-Garcia:2020cdo}, the commuting SYK of \cite{Gao:2023gta} (whose double scaling limit was studied in \cite{Almheiri:2024ayc,Berkooz:2024ofm}), models of interacting Paulis such as \cite{Watanabe:2023vzo} and the more general polarized spin operators of \cite{Berkooz:2024ofm}. Here we are going to highlight additional models which demonstrate some peculiar features, relevant for holography or for quantum chaos.

\subsection{The Parisi hypercube model}\label{sec:parisi}
\label{subsec:UnivClass}

 The first example is the Parisi hypercube model, introduced and solved in \cite{Parisi:1994jg} and recently studied in \cite{Jia_2020}. It is a case where chord diagrams are applied to random Hamiltonians that go beyond $p$-locality (a partial characterization of this class was suggested in \cite{Berkooz:2023scv,Berkooz:2023cqc}), and presents a new take on the role of frustration in chord models. One can use this approach of frustration to construct models with bosons   \cite{Jia:2024tii}, which are otherwise difficult to control.
 
In the conventions of  \cite{Berkooz:2023scv,Berkooz:2023cqc}, the hypercube has $2^d$ vertices denoted by $\{-\tfrac{1}{2},+\tfrac{1}{2}\}^d$ with $\frac{\sigma^3_\mu}{2}$ being the position operators of the particle, $\mu=1,\cdots,d$. The Hamiltonian is
\begin{equation}\label{eqn:symmetricGauge}
    H =- \frac{1}{\sqrt{d}}\sum_{\mu=1}^d  D_\mu \,, \quad D_\mu = T_\mu^+ + T_\mu^-, \quad T^+_\mu = \prod_{\nu=1, \nu\not=\mu}^d e^{\frac i4 F_{\mu\nu}\sigma^3_\nu} \sigma^+_\mu \,, \quad q\equiv  \langle \cos F_{\mu\nu} \rangle \,,
\end{equation}
where  $T^-_\mu = (T^+_\mu)^\dagger$  and $\sigma^+_\mu =\frac{\sigma^1_\mu + i \sigma^2_\mu}{2}$.
$\sigma_\mu^i$ ($i=1, 2, 3$)  is the $i$-th Pauli matrix acting on the $\mu$-th qubit, and $F_{\mu\nu}$ is the antisymmetric tensor of the  background flux. $T^+_\mu$ is a hopping operator that transports the particle in the forward $\mu$ direction, while assigning to it a random phase due to a disordered flux.  We have chosen a normalization for  $H$ such that it  has a compact spectral support at  $d = \infty$ (as in DS-SYK).  The fluxes  $F_{\mu\nu}$ are quench-disordered, and identically and independently distributed  with an even distribution, but otherwise completely general. $\langle \cdots \rangle$ Denotes ensemble average, 
and $q$ is a tunable parameter.

The holonomy  around a plaquette in the $\mu\nu$ plane, $T^-_\nu  T^-_\mu  T^+_\nu  T^+_\mu$, gives the return amplitude of hopping counter-clockwise around it. It is more convenient to study the holonomy in terms of the $D_\mu$ operator, which combines the forward and backward hoppings:
\begin{equation}\label{eq:frust}
      {\cal W}_{\mu\nu} = D_\nu D_\mu D_\nu D_\mu=\cos(F_{\mu\nu}) - i \sin(F_{\mu\nu}) \sigma^3_\mu  \sigma^3_\nu, \qquad 
      \langle {\cal W}_{\mu\nu} \rangle = q.
\end{equation}
We can also think of ${\cal W}_{\mu\nu}$ as the mutual frustration of different terms in the Hamiltonian.

Consider now computing the moments $
   \langle \Tr (H^{2k})\rangle = d^{-k} \sum_{\{\mu_i\}}  \langle \Tr
  (D_{\mu_1}  D_{\mu_2}\cdots  D_{\mu_{2k}})
    \rangle$ in the position basis. In the large $d$ limit, if a direction $\mu$ appears in the trace, then it appears twice. This gives us the basic chord diagram structure. If two chords cross it means that these four hopping operators appear in the ordering \eqref{eq:frust}, i.e. a chord intersection gives a factor of $q$, and we obtain the chord rules right away.

\subsection{Models with additional symmetries}

There are some classic generalizations of the SYK models (actually, some predate it) which are useful for experiments (the $U(1)$ invariant model) or for holography (the SUSY models). From the chord perspective they are interesting because the basic chord Hilbert space is seemingly a multi-chord, with seemingly a very different set of intersection rules, yet it can be reduced to a single chord Hilbert space. Beyond their own sake, it is an interesting problem to explore the relation between theories that can be reduced a simple transfer matrix and requirements of locality in the bulk.

The $U(1)$ invariant SYK model \cite{Davison:2016ngz,Sachdev:2015efa,Gu:2019jub}, was analyzed in the double scaling limit in \cite{Berkooz:2020uly}. The model contains $N$ Dirac fermions, $\psi^i$, and their complex conjugates ${\bar \psi}_i$
\begin{equation}
H=\sum_{I,K} J_I^K  {\bar\psi}_K  \psi^I \,,
\end{equation}
where $I,K$ are multi-indices (whose length we double scale). The $J$'s are complex Gaussian variable $J_I^K=(J_K^I)^*$, and as before their Wick contraction gives rise to a chords description. The weights of the chord diagrams appear differently -- the intersection of chords now has weight 1, and non intersecting chords carry non-trivial weights that depend on the chemical potential. Naively it seems that we cannot apply the transfer matrix technique. However with some effort one can reduce them to the same transfer matrix as above in a fixed charge sector. It is generally not clear what are the total allowed set of pairwise chord interactions and which ones can reduced to an ordinary transfer matrix. 

The situation is more interesting in the supersymmetric SYK model \cite{Fu:2016vas} which was solved in the double scaling in \cite{Berkooz:2020xne}. The model is 
\begin{equation}
Q \propto \sum_{I} C_I \Psi_I,\ \ \ \ {\cal N}=1:\ C,\psi\ real;\ \ {\cal N}=2:\ C,\psi\ complex
\end{equation}
The model with ${\cal N}=1$ is solved in terms of chords stretching between insertions of $Q$. This means that $H_{effective}=T^2(q_f)$ where $q_f$ is negative $q_f=-e^{-2p^2/N}$, and the Arik-Coon algebra \eqref{eq:ArikCoon} is then a deformation of anti-commutation relations. The case of ${\cal N}=2$ is more interesting. Since $Q^2=(Q^\dagger)^2=0$, it is enough to consider $\Tr(QQ^\dagger\cdots QQ^\dagger)$. At each stage either $Q$ or $Q^\dagger$ can annihilate or create a chord. Since the $C$'s are complex variables, the chords have an orientation and we will take it to go from a $Q$ to a $Q^\dagger$. When we cut the trace to go the transfer matrix we need to keep track of whether a chord goes left or goes right, so there are two species of chords, making the transfer matrix considerably more complicated, but one can reduce it to the ordinary transfer matrix via decoupling a vast number of null states in the chord Hilbert space. 

This model was used in \cite{Boruch:2023bte} to relate the fraction of ground states in the full Hilbert space to the wavefunctions on the chord Hilbert space. In the appropriate limit, it reproduces the Schwarzian result, while showing a series of non-perturbative corrections to it whose interpretation is not yet understood.

\subsection{Multi chord problems}\label{sec:MltChrd}

The generalization to multi-chord situation, i.e. when there are several types of chords with non-trivial intersection rules is quite ubiquitous in problems with RG flow. For example if we take  $K$  independent random $p$-local Hamiltonians on the same set of qubits with the total Hamiltonian being $H = \sum_{i=1}^K \kappa_i H_i$, where $\kappa_i$ are the relative sizes of the couplings. These Hamiltonians can be SYK-like operators of different lengths, or preserve additional symmetries as in the operator
\begin{equation}
    \sum_{\substack{\ \ \ 1 \le j_{1} < \cdots < j_{k^\prime} \le M \\ 2M < i_1 < \cdots < i_k \le N}} \tilde J_{i_1,\cdots,i_k}^{j_{1},\cdots,j_{k^\prime}} X_{j_{1}} \cdots X_{j_{k^\prime}} \psi_{i_1} \cdots \psi_{i_k} \,, \qquad X_j = \psi_{2j-1} \psi_{2j} ,
\end{equation}
for some $M \le N/2$, or have some non-uniform distribution of the random couplings $J$. For various realizations, see Appendix B in \cite{Berkooz:2024ofm}.

Each Hamiltonian has its own species of chords associated with it, originating from the Wick contraction of its couplings. We assign a weight $0\le q_{ij}\le 1$ \cite{Berkooz:2024ofm} to the intersection of chords of types $i$ and $j$ (including $i=j$).


One can again define a chord Hilbert space and transfer matrix by introducing $q$-deformed chord creation and annihilation operators \cite{Speicher:1993kt, pluma2022dynamical} and a transfer matrix,
\begin{equation}
    [a_i, a_j^\dagger]_{q_{ij}} = \delta_{ij} \,,\qquad T = \sum_{i=1}^K \kappa_i \left(a_i + a_i^\dagger\right) \,, \qquad Z(\beta) = \langle \vec 0 | e^{-\beta T} |\vec 0 \rangle \,.
\end{equation}
Unfortunately, if one tries to use the transfer matrix method then one needs to track not just the total number of chords, but rather the number of chords for each of the species as well as their ordering, making the Hilbert space exponentially more complicated and rendering it impractical to work with (as of yet) -- even the size of the sub-space with $n$ chords in total is $K^n$. Very little is known about the transfer matrix in this challenging case.

Nevertheless, one can write a  ``path intergal'' formula for chords, which simplifies considerably in the semi-classical limit $q_{ij}\to 1$ \cite{Berkooz:2024ofm, Berkooz:2024evs}. At the core of the approach one splits the thermal circle into small intervals and uses variables that count how many chords of different types go between all pairs of intervals -- i.e. the density of chords between two points on the thermal circle, $n_i(\tau_1,\tau_2)$. The expression\footnote{The expression in terms of the chord density only agrees with this one on-shell, and in fluctuations around the saddle point. This issue has not been fully resolved yet.} looks especially suggestive when expressed in terms of the variables\footnote{When $K=1$ the variable $g(\tau_1,\tau_2)$ is proportional to length of bulk geodesics that connect these two boundary points, and therefore $n(\tau_1,\tau_2) \sim \partial_1 \partial_2 g(\tau_1,\tau_2)$ connects the chord density to the Crofton form -- the natural measure on the space of bulk geodesics, see \cite{Czech:2016xec}.}, $g_i(\tau_1, \tau_2) \sim \int_{\tau_1}^{\tau_2} d\tau_3 \int_{\tau_2}^{\tau_1}d\tau_4 \, n_i(\tau_3,\tau_4)$, which counts how many chords go across a probe chord that would stretch between the points $\tau_1$ and $\tau_2$,
\begin{equation}\label{eq:Mult}
    Z = \int \left(\prod_{i=1}^K \cD g_{i}\right) \exp\left[ -\frac{1}{\lambda}\int_0^\beta d\tau_{1}\int_0^{\tau_1}d\tau_{2}\left[\sum_{i,j=1}^K\frac{1}{4}\alpha_{ij}g_{i}\partial_{1}\partial_{2}g_{j} + \sum_{i=1}^K\kappa_{i}^2 \bJ^2 e^{\sum_{j}\alpha_{ij}g_{j}}\right] \right] \,,
\end{equation}
where $q_{ij}=e^{-\lambda_{ij}}$, $\alpha_{ij} = \lambda_{ij} / \lambda$, and $\lambda$ is some parameter that uniformizes the $q_{ij}\to 1$ limit. This action\footnote{We thank Josef Seitz and Ronny Frumkin for discussions of this point.} generalizes the $G\Sigma$ form \eqref{eq:GSigma double scaling} for these multi-chord diagrams.

A particularly interesting case \cite{Berkooz:2024ofm, Berkooz:2024evs} arises when $K = 2$ and the Hamiltonian of the system has a chaotic part and an integrable part\footnote{Such double scaled Hamiltonians where $q=1$ were recently studied in \cite{Almheiri:2024xtw}.}, 
\begin{equation}H = \kappa_1 H_{chaotic} + \kappa_2 H_{integ} \,.
\end{equation} 
For example, $H_{chaotic}$ could be the $p$-spin Hamiltonian \eqref{eq:SpinHamilt} or the SYK Hamiltonian \eqref{eq:SYK Hamiltonian} and $H_{integ}$ could be $\sum_I J_I \sigma^z_{i_1}...\sigma^z_{i_{p'}}$ or the ``commuting SYK'' model \cite{Gao:2023gta}, respectively. In these cases $q_{11}$ and $q_{12}$ are arbitrary ($<1$) and $q_{22} = 1$. Using the above technique one can map the phase diagram of the chaos to integrability transition and establish that it is first order at low temperatures (perhaps with implications for the experimental setup \cite{Jafferis:2022crx}). 

Other Hamiltonians were also analyzed \cite{Berkooz:2024ofm} -- a mixture of SYK Hamiltonians does not exhibit a phase transition \cite{Anninos:2022qgy}, while other choices of $q_{ij}$'s give a zero temperature phase transition.


\section{Some additional frontiers}\label{sec:Frntrs}

Finally, we're going to briefly scan through some current topics of research (in addition to some mentioned in the sections before), restricted to a level of very rough pointers.

\subsection{The bulk}

We have seen a few time the appearance of unusual "bulks". One of them had to do with {\bf Non-commutative geometry} which is playing an important role in this model, as it keeps showing up, in different guises, at almost every step. At the same time, it is evident from Section \ref{sec:non-commutative} that there is no complete story yet. The maximal goal would be a constructive derivation of the full 2D non-commutative geometry (or integral over matrices, or some generalization of the latter) directly from the microscopics, without any answer analysis. 

The semi-classical {\bf fake geometries} are another set of puzzling geometries which emerges from the discussion. At face value one excises pieces of the geometry for no apparent reason, and it is not clear (at least to us) how to use them for higher $n$-pt functions. It might have to do with the fact the inner product spreads across a finite volume of spacetime rather then being localized on a Cauchy surface.

Some other issues which we have not discussed include:

\paragraph{A gravitational path integral}
\label{sec:Gravitational path integral}

A natural question one might ask is whether the DS-SYK model can be recast in the form of a gravitational path integral, where one has to integrate over some bulk metrics of usual (commuting) spacetimes. Such a path integral formulation was suggested in \cite{Verlindetalk, Goel:2022pcu, Blommaert:2023opb} and recently analyzed in \cite{Blommaert:2024ydx}. It takes the form of the sine-dilaton theory
\begin{equation}
    \int \cD g \cD\Phi \exp\left[\frac{1}{2\lambda} \int_{\cM} d^2x \sqrt{g} \left(\Phi R + 2\sin(\Phi)\right) + \frac{1}{\lambda}\int_{\partial\cM} d\tau \sqrt{h} \left( \Phi K - e^{-\frac{i\Phi}{2}} \right) \right] \,,
\end{equation}
supplemented by the boundary conditions $\Phi\big|_{\partial\cM} = \frac{\pi}{2} + i\infty$, $\sqrt{h}e^{-\frac{i\Phi}{2}}\big|_{\partial\cM} = \frac{i}{\lambda}$. The theory was canonically quantized in \cite{Blommaert:2024ydx}, analogously to JT-gravity \cite{Harlow:2018tqv}, by observing that a natural canonical coordinate is $L = \int ds e^{-\frac{i\Phi}{2}}$, the worldline action of a particle with non-minimal coupling to the dilaton. The Hamiltonian of this system then becomes the transfer matrix \eqref{eq:q-Liouville Hamiltonian} as a finite difference operator, but additional steps are needed to get the correct chord Hilbert space. When $\lambda \to 0$, these steps include inserting a superposition of defects in the bulk, somewhat resembling the fake disk picture. The resolution for finite $\lambda$ is unclear. 

This proposal also suggests a first order formulation (which generalizes JT's BF gauge theory formulation) in terms of a Poisson-sigma model \cite{Blommaert:2023opb, Blommaert:2023wad} -- a non-linear sigma model whose target space is the phase space of a quantum mechanical system -- which manifests the quantum group symmetry, and implies the particle on a group description of Section~\ref{sec:non-commutative}.

\paragraph{Yang-Baxter moves and geometry}

\begin{figure}[h]
\begin{subfigure}[t]{0.45\textwidth}
\centering
\includegraphics[width=0.4\textwidth]{"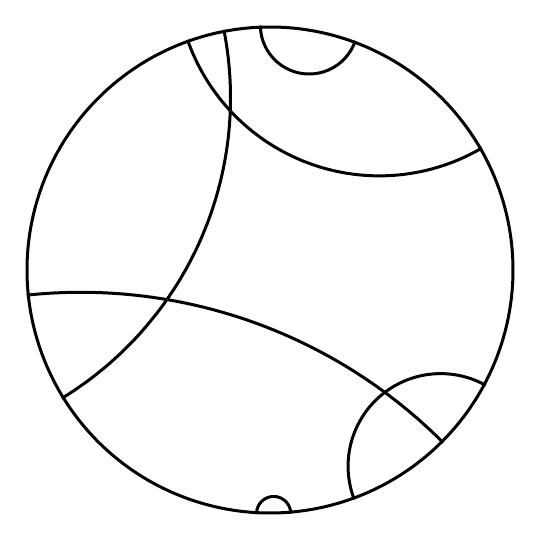"}
\hfill
\includegraphics[width=0.4\textwidth]{"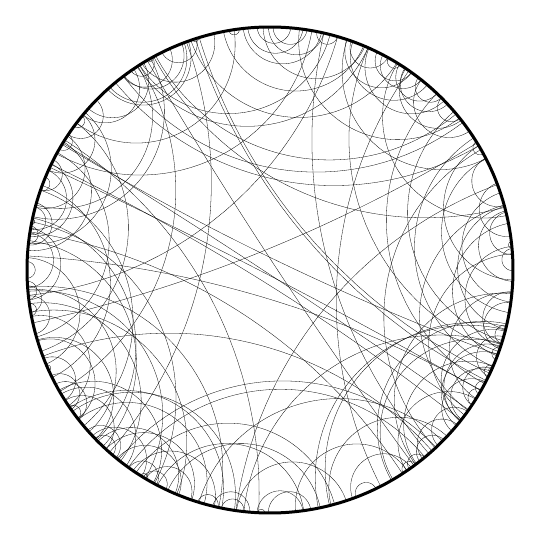"}
\caption{}
\label{ChordIntro}    
\end{subfigure}
\qquad
\begin{subfigure}[t]{0.45\textwidth}
\centering
    \includegraphics[width=0.9\textwidth]{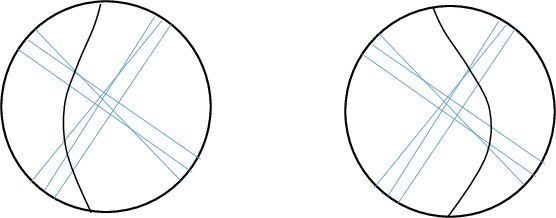}
    \caption{}
    \label{fig:YBmoves}    
\end{subfigure}
\caption{(a) Chord diagrams proliferating at $q\rightarrow 1$, (b) The same chord diagram but different bulk realizations.}
\end{figure}

The relation between the chord picture and the bulk is confusing at a more fundamental level. In the limit $q\rightarrow 1$ chord crossings are not suppressed and they fill the disk, Figure \ref{ChordIntro}. So a 2D bulk seems to arise naturally, and we would like to have some sort of a map from chord diagrams to metrics (for example).

But making this precise is tricky. Chord diagrams only provide information on which chords intersect, and no information on how they traverse the bulk. In fact we can move them around by ``Yang-Baxter moves". For example in \autoref{fig:YBmoves}
the black chord has the same intersection numbers with the blue chords but it runs in a different location relative to the blue mutual intersection. If we believe that how chords traverse the bulk of the disk teaches us about the bulk geometry, then these could be seen as different bulk geometries. So even though \autoref{ChordIntro} is suggestive, we cannot yet construct the bulk directly.

The main reason for obsessing about this is that a key paradigm in string theory and field theory, due to `t Hooft, is that 2D surfaces are intimately related to the double line large $N$ expansion. In the language of chords these would be non-intersecting chords but with higher order vertices (determined by the model's potential). Here we have a different\footnote{There were attempts at reproducing the density of states and correlators of DS-SYK using a 2-matrix model \cite{Jafferis:2022wez} (see also \cite{Okuyama:2023kdo,Okuyama:2023yat}). For example, formally one can take a multi-matrix ensemble and integrate it over orbits of the unitary group. However, DS-SYK models are significantly different from ordinary RMT ensembles in other features such as multi-trace correlators where they have much larger early time connected contributions.} way of obtaining 2D surfaces.

\subsection{Level statistics, multi-trace correlators, and wormholes}
\label{sec: wormholes}

In this review we have discussed the strict double scaling limit, where $N = \infty$, $\lambda$ is kept fixed, and the spectrum is continuous, as discussed at the end of Section~\ref{sec:TrnsfrMtrx}. We had also concentrated on single trace quantities, which measure the ensemble averages of observables. But we might also be interested in the statistics and correlations of these observables within the ensemble. These are measured by multi-trace quantities, and go beyond the strict $N = \infty$ limit. 

One such key observable in any random systems is the Spectral Form Factor (SFF) $\langle \Tr(e^{-(\beta+it)H }) \Tr(e^{-(\beta-it)H}) \rangle_J $, which has in principle the same information as 2-energy level statistics.
For a Wigner-Dyson ensemble, the SFF is made out of a ramp and a plateau. The former is affected by the spectral rigidity of the model -- up to which scale one can see RMT statistics in the spectrum. The latter is affected by the ultimate discreteness of the spectrum. For a discussion in the context of SYK see \cite{Cotler:2016fpe, Saad:2018bqo, Gharibyan:2018jrp} and \cite{Saad:2019lba, Stanford:2019vob} for JT gravity. This observable is just the simplest out all possible multi trace correlators $\langle Z(\beta_1)\cdots Z(\beta_k)\rangle_J$. In holography these are related to multiboundary Euclidean wormholes. 

In these computations we can reinstate the $N$ dependence - for example in the trace, or in the number of independent couplings -- and obtains results that highly suppressed in $N$, but are leading connected contributions to the multitrace correlators. This is similar to taking into account the $1/\dim({\cal H})$ suppression in the topological recursion relations in random matrix theory.  

However, in the full SYK model, as in any realistic model beyond the strict Wigner-Dyson ensembles, there are much larger model dependent contributions at shorter time scales, and to a large extent they are terra incognita. (Un)Fortunately these are the time scales where the OTOC, Hawking radiation and other interesting effects take place. These contributions were evaluated in SYK or DS-SYK (the results are similar, but may differ for other chord models) -- their size in \cite{Altland:2017eao} and their exact form in \cite{Berkooz:2020fvm}. These effects scale as $1/N^p$ -- much larger than the ramp and the plateau, which are exponentially small in $N$. They correspond to global fluctuations of the spectrum which shift its entire shape, and not just nearby energies (for the implications to edge fluctuation see \cite{Altland:2024ubs}). Ultimately, at very late times, we expect to see the ramp and plateau behavior, as we are dealing with chaotic systems with discrete spectrum. Yet, in these double scaled models (or, at the very least, in SYK) there seem to be other important effects at earlier timescales.

Practically, this result is also relevant for sample-to-sample observable quantities in potential experimental realization of the SYK system \cite{Shackleton:2023lpw}. 
From the point of view of gravity this is puzzling -- for example, two boundaries of size $\beta_1$ and $\beta_2$ ``communicate" with each other via effects which are much larger than what the ``double trumpet" geometry predicts. It is an open problem how to fully understand them in gravity \cite{Berkooz:2020fvm}.

\subsection{Relation to (operator) complexity}
\label{sec:complexity}

A useful way of characterizing quantum chaos is by operator growth \cite{Roberts:2018mnp,Parker:2018yvk}. There are several possibilities how to parameterize this growth, with Kyrlov complexity being one popular method \cite{Parker:2018yvk,Barbon:2019wsy}. The starting operator can also be different. One option is to start with the identity operator and evolve it by $e^{-{(\beta+it)} H}$ -- this will measure\footnote{also, it describes how the complexity of the identity operator increases as we project it to subspaces of lower energy using $e^{-\beta H}$.} the complexity of the thermofield double state\footnote{Regarding the space of operators as the doubled Hilbert space.}, and correspondingly the two sided wormhole, or its (upward) time evolution on both sides. Another starting point could be a simple enough operator, and then we evolve the wormhole with a single particle in it. 

In the case of SYK the growth is parameterized in terms of the number of fermions in the operators, i.e., by the super-operator $\hat n(\cO) = \frac{1}{4p}\sum_{k=1}^\infty [\psi_k,[\psi_k,\cO]]$ where $\cO$ is the operator whose length we want to measure \cite{Qi:2018bje}. In DS-SYK this length is just the number of chords: if we have $\hat n$ chords open, then the number of fermions in the state is ${\hat n} p$. Microscopically, there are some fermions in the overlap of the index sets of chords but they are of order ${\hat n}^2 \ll {\hat n} p$ in the time regime where the chord method applies. For the TFD state, the chord number measures the Krylov state complexity \cite{Lin:2022rbf,Rabinovici:2023yex}. We've seen before that $\hat n$ is also the length of the wormhole connecting the two sides of the TFD state, and therefore it provides a concrete realization of the idea that complexity is encoded in the bulk geometry (\cite{Susskind:2014rva, Brown:2015bva, Stanford:2014jda} and many others). The length also grows for simple operators\footnote{For an attempt to relate that to a precise notion of complexity see \cite{Aguilar-Gutierrez:2024nau}.} \eqref{eq:finite q eom}, and for other models which have a chord description. Specifically for the case of DS-SYK one can actually go further and map the operator growth into motion in a Parisi hypercube model \cite{Berkooz:2023scv,Berkooz:2023cqc}.

\subsection{de Sitter space}
\label{sec:de Sitter}

Some features of DS-SYK had prompted conjectures that at high temperatures it describes the physics of quantum gravity on de Sitter space \cite{Susskind:2021esx,Susskind:2022dfz,Lin:2022nss,Susskind:2022bia,Rahman:2022jsf,Susskind:2023hnj,Rahman:2023pgt,Rahman:2024vyg,Narovlansky:2023lfz,Verlinde:2024znh,Verlinde:2024zrh,Verlindetalk}.

First, its spectrum is bounded, as we have seen in Section~\ref{sec:HamChords}. As such, the (one-sided) Hamiltonians do not diverge in the large $N$ limit. In addition, there is a maximally mixed state with finite energy, and  equivalently, a maximally entangled state on the doubled Hilbert space with finite energy, which is the infinite temperature TFD state. In a related fashion, the Hamiltonian and matter operators of DS-SYK generate a type II$_1$ von-Neumann operator algebra\footnote{The connection between the $q$-oscillator algebra \eqref{eq:ArikCoon} and type II$_1$ operator algebras was long known in the mathematics community, for example \cite{speicher1997q, ricard2003factoriality}.} \cite{Lin:2022nss}, see also \cite{Xu:2024hoc}. This is exactly the type of algebra of observables in the static patch of de Sitter space \cite{Chandrasekaran:2022cip}.

Second, despite being at infinite temperature, there are still some notions of finite temperature in the system. One is the fake temperature of Section~\ref{sec:semi-classics}, that controls the decay of 2-pt functions. Another is the tomperature \cite{Lin:2022nss}, that measures the change in the energy of the system as the number of fermions changes. Both have analogs in $dS_2$ \cite{Rahman:2024vyg}, prompting a proposed duality between the $\lambda\to 0$, infinite temperature limit of DS-SYK and JT $dS_2$ \cite{Susskind:2022bia, Rahman:2022jsf}. These proposals rely on the semi-classical features of the models. More holographic properties at high temperatures were investigated in \cite{Almheiri:2024xtw}.

Another proposal relates DS-SYK to pure gravity in three-dimensional de-Sitter space, $dS_3$ \cite{Verlindetalk,Verlinde:2024znh,Verlinde:2024zrh,Narovlansky:2023lfz}. Interpreting two copies of the DS-SYK model (with a constraint) as the worldlines of particles in $dS_3$, one can match between correlation functions in the former and (S-wave) scalar Green's functions in the latter in the high-temperature, semi-classical limit \cite{Narovlansky:2023lfz}. But the proposal also extends to the quantum regime. The phase-space of Schwarzschild-$dS_3$ solutions for pure gravity can be quantized. The Hamiltonian is a certain line operator in the theory, which was shown to have the same form as the DS-SYK transfer matrix \eqref{eq:q-Liouville Hamiltonian} \cite{Verlinde:2024zrh, Verlinde:2024znh}, where $\lambda$ corresponds to the ratio between the Planck scale and the de Sitter radius. 

Another evidence for the relation comes from treating pure gravity on dS$_3$ as an $SL(2,\bC)$ Chern-Simons theory, so the quantization described above is a proposed quantization of the latter. The wavefunctions of states in its Hilbert space are related to correlation functions in a system of two 2d Liouville CFTs with conjugate central charges, $c_\pm = 13 \pm 12\pi i(\frac{\lambda}{4\pi^2} - \frac{1}{\lambda})$, termed the the dS-Liouville model. The boundary two point functions in this dS-Liouville model were then shown to agree with the two-point functions of the doubled DS-SYK model mentioned above for any $\lambda$ \cite{Verlinde:2024znh,Verlinde:2024zrh}.

It would be interesting to have a clearer picture of the relation between the proposals, and to understand the role of the sine-dilaton gravity, given its relation to the Liouville theory mentioned above, and the existence of classical saddles in the theory where some regions in the bulk have positive curvature \cite{Blommaert:2023opb}.

\paragraph{Acknowledgments}
We would like to thank A. Almheiri, A. Altland, A. Blommaert, N. Brukner, R. Frumkin, F. Haehl, M. Isachenkov, Y. Jia, P. Narayan, V. Narovlansky, A. Raz, S. Ross, M. Rozali, Josef Seitz, J. Simon and M. Watababe for many useful discussions. OM is supported by the ERC-COG grant NP-QFT No. 864583
``Non-perturbative dynamics of quantum fields: from new deconfined
phases of matter to quantum black holes", by the MUR-FARE2020 grant
No. R20E8NR3HX ``The Emergence of Quantum Gravity from Strong Coupling
Dynamics", and by the INFN ``Iniziativa Specifica GAST''. The work of MB is supported in part by the Israel Science Foundation grant no. 2159/22, by the Minerva foundation, and by a German-Israeli Project Cooperation (DIP) grant "Holography and the Swampland".

\appendix

\section{Some $q$-functions}\label{app:MathDef}
Here we define the special functions used throughout this work. We will use $q = e^{-\lambda}$. 

\paragraph{The $q$-Pochhammer symbol}
The $q$-Pochhammer symbol is defined to be
\begin{equation}
    \left(z;q\right)_{n} = \prod_{k=0}^{n-1}\left(1-zq^{k}\right) \,,
\end{equation}
and the shorthand notation $(a_1,\cdots,a_k;q)_n = \prod_{i=1}^k (a_i;q)_n$ and $(e^{\pm i\theta};q)_n = (e^{i\theta};q)_n (e^{-i\theta};q)_n$. 

The infinite $q$-Pochhammer symbol has the expansions
\begin{equation}
\label{eq:Pochhammer plethystic}
\left(z;q\right)_{\infty} = \sum_{n=0}^{\infty}\frac{\left(-1\right)^{n}q^{\frac{n\left(n-1\right)}{2}}}{\left(q;q\right)_{n}}z^{n} = \exp\left[-\sum_{k=1}^{\infty}\frac{1}{k}\frac{z^{k}}{1-q^{k}}\right] \,.
\end{equation}

\paragraph{The Jacobi theta function}
The Jacobi theta functions are quasi-elliptic functions of two variables. We will use the conventions of \cite{Kharchev_2015}, 
\begin{equation}
\label{eq:Jacobi_theta_series}
    \theta_{1}\left(u|\tau\right) = -i\sum_{k=-\infty}^\infty(-1)^k \fq^{(k+\frac{1}{2})^2} e^{\pi i(2k+1)u} = 2\fq^{1/4}\sin\left(\pi u\right)\left(\fq^2,\fq^2e^{\pm2\pi iu};\fq^2\right)_{\infty} \,,
\end{equation}
where $\fq = e^{\pi i\tau}$. The theta function has the modular property
\begin{equation}
\label{eq:Jacobi_theta_modular}
    \theta_1\left(u | \tau\right) = \frac{i}{\sqrt{-i\tau}} e^{-\pi i u^2 / \tau} \theta_1\left(\frac{u}{\tau} \,\bigg| -\frac{1}{\tau}\right) \,.
\end{equation}

\paragraph{$q$-numbers}
We will sometimes use the definition of a $q$-number,
\begin{equation}
    [n]_q \equiv \frac{1-q^n}{1-q} = 1 + q + \cdots + q^{n-1} \,,
\end{equation}
where the last equality is valid for integers. When $q\to 1$ and $n$ is fixed, $[n]_q \to n$.

\paragraph{The $q$-factorial and $q$-Gamma function}
The $q$-Pochhammer symbol can also be used to define the $q$-factorial and $q$-Gamma function, 
\begin{equation}
\label{eq:q factorial def}
    \Gamma_{q}\left(n\right) \equiv \left[n-1\right]_{q}! \equiv \prod_{k=1}^{n-1} [n]_q = \frac{\left(q;q\right)_{\infty}}{\left(q^{n};q\right)_{\infty}}\frac{1}{\left(1-q\right)^{n-1}} = \frac{(q,q)_{n-1}}{(1-q)^{n-1}}\,.
\end{equation}
In the limit $q\to 1$ the $q$-Gamma function becomes the usual Gamma function, $\lim_{q\to 1} \Gamma_q(x) = \Gamma(x)$ when $x$ is independent of $q$. 

\paragraph{The $q$-exponential}
The $q$-deformation of the exponential function is\footnote{In another standard convention where  $e_a(z;\hq^2) \colonequals \sum_{n=0}^\infty { z^n \over \hq^{an(n-1) \over 2} (\hq^2 ;\hq^2)_n  }$ it is $e_0$.}
\begin{equation}
    \label{eq:q-exponential def}
    e_q(z) = e_q^z = \sum_{n=0}^\infty \frac{z^n}{[n]_q!} = \sum_{n=0}^\infty \frac{z^n(1-q)^n}{(q;q)_n} \,, \qquad e_q\left(\frac{z}{1-q}\right) \xrightarrow{q\to 1} e^z \,.
\end{equation}
It has a nice property with respect to the $q$-derivative, $D_q e_q^z={e_q(qz)-e_q(z)\over qz-z}=e_q(z)$.

\paragraph{The continuous $q$-Hermite polynomials}
The continuous $q$-Hermite polynomials are a system of orthogonal polynomials. Their generating function is
\begin{equation}
    \sum_{n=0}^\infty \frac{H_n(\cos(\theta)|q)}{(q;q)_n} t^n = \frac{1}{(te^{i\theta};q)_\infty (te^{-i\theta};q)_\infty} \,.
\end{equation}
A useful form of these polynomials is \cite{ismail1987combinatorics}
\begin{equation}
    H_n(\cos\theta|q) \equiv \sum_{k=0}^n \binom{n}{k}_q e^{i(n-2k)\theta} = \sum_{k=0}^n \frac{(q;q)_n}{(q;q)_{n-k}(q;q)_k} e^{i(n-2k)\theta} \,.
\end{equation}
The $q$-Hermite polynomials satisfy the recursion relation
\begin{equation}
    2x H_n(x|q) = H_{n+1}(x|q) + (1-q^n)H_{n-1}(x|q) \,,\qquad H_{-1}(x|q) = 0 \,, \quad H_0(x|q) = 1\,.
\end{equation}
They are orthogonal polynomials in $\theta$,
\begin{equation}
    \label{eq:q_hermite_theta_ortho}
    \int \frac{d\theta}{2\pi} (q,e^{\pm 2i\theta};q)_\infty H_m(\cos \theta|q) H_n(\cos\theta|q) = (q;q)_n \delta_{nm} \,, 
\end{equation}
and satisfy the relation
\begin{equation}
    \label{eq:q_hermite_n_orthogonality}
    \sum_{n=0}^\infty H_n(\cos \theta|q) H_n(\cos \phi|q) \frac{t^n}{(q;q)_n} = \frac{(t^2;q)_\infty}{(te^{i(\pm\theta\pm\phi)};q)_\infty} \,.
\end{equation}

\paragraph{The $R$-matrix}
The $6j$-symbol of $\cU_{\sqrt{q}}\left(\fsu(1,1)\right)$ takes the explicit form
\begin{multline} \label{eq:R_matrix_full_expr}
R^{(q)} _{\theta _4\theta _2} \begin{bmatrix}\theta _3 & l_2 \\ \theta _1 & l_1 \end{bmatrix} 
= \frac{\left( \tilde q_1 e^{-i(\theta _2+\theta _3)} ,\tilde q_1 \tilde q_2 e^{i(\theta _3 \pm \theta _1)} ,\tilde q_1\tilde q_2e^{i(\theta _2 \pm \theta _4)} ;q\right)_{\infty } }{\left( \tilde q_1 \tilde q_2^2 e^{i(\theta _2+\theta _3)} ;q\right)_{\infty } }  \\
\times \frac{(\tilde q_2^2;q)_{\infty } }{\sqrt{\left( \tilde q_1 e^{i(\pm \theta _2 \pm \theta _3)} ,\tilde q_1 e^{i(\pm \theta _1 \pm \theta _4)} ,\tilde q_2 e^{i(\pm \theta _1 \pm \theta _2)} ,\tilde q_2 e^{i(\pm \theta _3 \pm \theta _4)} ;q\right)_{\infty } } }  \\
\times {}_8W_7\left( \frac{\tilde q_1\tilde q_2^2 e^{i(\theta _2+\theta _3)} }{q} ;\tilde q_1 e^{i(\theta _2+\theta _3)} ,\tilde q_2e^{i(\theta _2\pm \theta _1)} ,\tilde q_2 e^{i(\theta _3 \pm \theta _4)} ;q,\tilde q_1e^{-i(\theta _2+\theta _3)} \right) \,,
\end{multline}
where ${}_8W_7$ is the basic one-variable well-poised hypergeometric series,
\begin{equation}
    {}_8W_7(a;b,c,d,e,f;q,z) = \sum_{n=0}^\infty \frac{\left(a,\pm qa^{1/2},b,c,d,e,f;q\right)_n}{\left(\pm a^{1/2},qa/b,qa/c,qa/d,qa/e,qa/f,q;q\right)_n} z^n \,.
\end{equation}

\printbibliography
\end{document}